\documentclass{article}

\usepackage{arxiv}

\usepackage[utf8]{inputenc} 
\usepackage[T1]{fontenc}    
\usepackage{hyperref}       
\usepackage{url}            
\usepackage{booktabs}       
\usepackage{amsfonts}       
\usepackage{nicefrac}       
\usepackage{microtype}      
\usepackage{lipsum}
\usepackage{graphicx}
\graphicspath{ {./images/} }
\usepackage{anyfontsize}
\usepackage{multirow}

\usepackage{textcomp, gensymb}
\usepackage{physics}
\usepackage{bm}
\usepackage{amsmath}
\usepackage{xcolor}
\usepackage{algorithm}
\usepackage{gensymb}
\usepackage{algpseudocode}

\title{Dynamic Behavior of Origami Structures: Computational and Experimental Study}

\author{
 Sudheendra Herkal   \\
  Department of Civil and Environmental Engineering\\
  Rice University \\
  Houston, TX 77054 \\
  \texttt{srh11@rice.edu} \\
   \And
 Satish Nagarajaiah* \\
  Department of Civil and Environmental Engineering and Department of Mechanical Engineering\\
  Rice University\\
  Houston, TX 77054 \\
  \texttt{satish.nagarajaiah@rice.edu} \\
  \And
 Glaucio Paulino \\
  Department of Civil and Environmental Engineering\\
  Princeton Materials Institute, Princeton University\\
  Princeton, New Jersey - 08544 \\
  \texttt{gpaulino@princeton.edu} \\
  * - Corresponding author
}
\begin{document}
\maketitle
\begin{abstract}
Origami structures have been receiving a lot of attention from engineering and scientific researchers owing to their unique properties such as deployability, multi-stability, negative stiffness, etc. However, dynamic properties of origami structures have not been explored much due to a lack of validated analytical dynamic modeling approaches. Given the range of interesting properties and applications of origami structures, it is important to study the dynamic behavior of origami structures. In this study, a dynamic modeling approach for origami structures is presented considering distributed mass modeling, which has the potential to be a generalizable approach. In the proposed approach, stiffness is modeled using the bar and hinge modeling approach while the mass is modeled using the mass distribution approach. Various candidate mass distribution approaches were investigated by comparing their responses to the finite element method responses for various geometric conditions, loading and boundary conditions, and deformation modes. It was observed that a dynamic modeling approach with triangle circumcenter mass distribution was able to capture most of the dynamics satisfactorily consistently. Subsequently, a Miura-ori specimen was manufactured and its free vibration response was determined experimentally and then compared to the prediction of the analytical model. The comparison demonstrated that the analytical model was able to capture most of the dynamics in the longitudinal direction.
\end{abstract}


\section{Introduction}
Origami is an ancient Japanese art form that has been receiving a lot of attention recently from engineering and scientific researchers. Origami structures consist of panels and folds, arrangement of which leads to a vast design space for origami structures. Many origami structures in this vast design space offer very interesting properties including some properties that cannot be found in traditional materials. For instance, some origami structures such as Miura-ori\cite{schenk2013geometry} or waterbomb\cite{ma2020folding} possess negative Poisson's ratio. Many origami structures have the property of being deployable, which is the ability to transform from folded state to unfolded state and vice versa multiple times, preferably with low actuation energy. Some of examples of deployable origami structures are the Kresling pattern \cite{zhang2018bistable} and the flasher pattern \cite{zirbel2013accommodating}. The deployability of the origami structures is used for applications where space constraints exist, such as for deploying solar panels and other systems in space. Recently, the James Webb Space Telescope was launched, wherein the folding and unfolding mechanisms were based on origami structures \cite{irion2010origami}.       
Apart from deployability, origami structures also offer the property of multi-stability. Multi-stability is the presence of multiple stable states for a given origami pattern. Many origami structures exhibit instability between these stable states, which results in them exhibiting negative stiffness \cite{li2019architected}. Some of the examples of origami structures that are multi-stable are the Kresling pattern \cite{novelino2020untethered}, curved origami structures\cite{zhai2020situ},  Water-bomb base structure \cite{hanna2014waterbomb} and Leaf-out pattern \cite{yasuda2016multitransformable}. This property of origami structures has been exploited for the actuation of robots \cite{novelino2020untethered,chi2022bistable} as they allow for swift and easy actuation as well as for vibration isolation \cite{sengupta2018harnessing,zhai2020situ}. It should be noted that most of these interesting properties are used for applications that are truly dynamic in nature.

The quasi-static behavior of origami structures is well understood, thanks to various quasi-static modeling approaches. The quasi-static behavior of origami structures is modeled either using analytical models, finite element models, or reduced order models. Analytical models are available for simple geometries and might be difficult to obtain if the geometry of the origami is complicated. The hinges in origami can pose a challenge for finite element modeling of origami and are modeled typically by manually connecting individual panels with hinge elements. Therefore, not only is finite element modeling (FEM) of origami structures computationally expensive, but it can be quite challenging to prepare the model as well. Hence, reduced order models have been proposed. The most widely used reduced order is the bar and hinge model \cite{filipov2017bar}. This model was further enhanced and implemented in MERLIN \cite{liu2017nonlinear} and MERLIN \cite{liu2018highly} softwares. MERLIN has been widely used for modeling the quasi-static behavior of various origami structures. 

Most of the deployment or morphing processes of origami structures are truly dynamic processes. Despite this, the dynamic behavior of origami structures unlike their static behavior, is not well understood as the dynamics of these structures is itself a nascent field. Some studies have modeled the dynamic behavior of origami materials analytically as well as numerically \cite{kidambi2019deployment, fang2017dynamics, xia2019dynamics}. However, most of the models are analytically derived for specific geometries and loading and boundary conditions or lack validation. Therefore, there is a need for validated dynamic models that can be used for modeling under varying geometric, loading, and boundary conditions. While it is possible to analyze the dynamic behavior of origami structures under varying conditions through FEM, the analysis of the dynamics further increases the computational cost. Therefore, it is beneficial if reduced model orders are capable of capturing the dynamics as well. Moreover, the distributed mass modeling of origami structures has not been studied, which can potentially lead to a more generalizable dynamic modeling approach. Given the range of interesting properties that origami structures display as well as the range of their applications, it is important to study the dynamic behavior of origami materials. In this study, the dynamic modeling of origami materials is presented considering the distributed mass modeling, which can be used to understand the dynamics of origami structures.

The organization of the paper is as follows. First, the formulation of the dynamic modeling for origami structures is described in the second section. The principal challenge with the formulation is the mass matrix, the derivation of which is extremely challenging as it is both time-varying as well as geometry dependent. Therefore, various mass distribution models were used in this study and are described in the third section. In order to empirically choose the best mass distribution model, various scenarios were used to test each of the mass distribution models, and the best model was determined by comparing the responses of the models to the response determined using FEM. This is presented in the fourth section. Experimental validation was performed to further validate the dynamic modeling approach and is presented in the fifth section. Finally, conclusions are presented in the last section.   

\section{Formulation of the dynamic equations} \label{sec:dyn_model}
\begin{figure}
 \centering
  \includegraphics[width=0.8\linewidth]{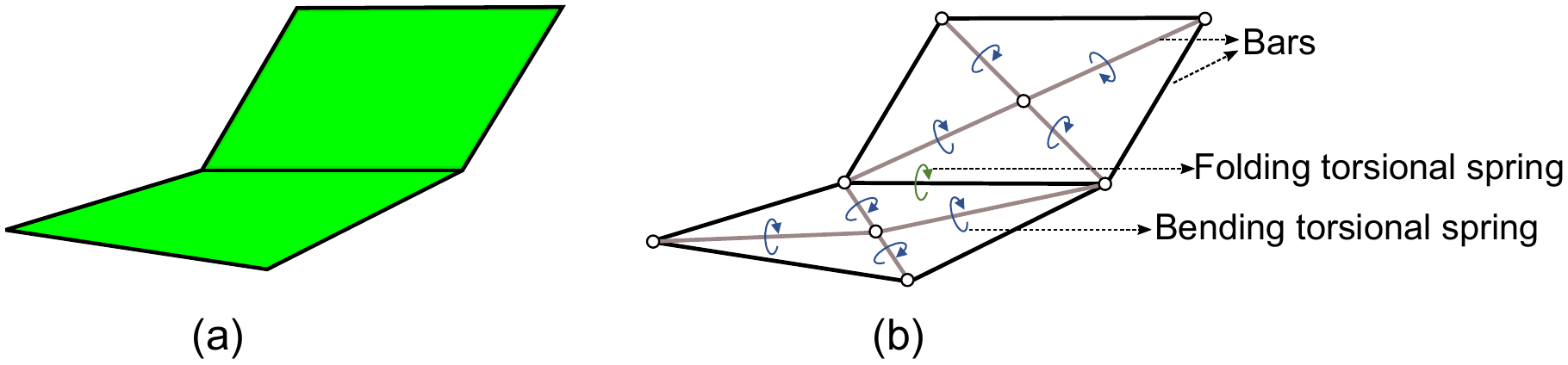}
  \caption{Discretization in MERLIN: (a) actual fold, and (b) discretized structure. }
  \label{fig:MERLIN}
\end{figure}

The formulation of dynamic equations is described in this section. When the origami structure is subjected to dynamic loads, the total energy of the origami structure consists of kinetic energy and potential energy parts and can be written as follows

   \begin{equation}
    L = K_{E} - P_{E} 
    \label{eqn:L}
\end{equation} 

where, $L$ is the Lagrangian energy function, $K_E$ is the kinetic energy and $P_E$ is the potential energy. In this study, the bar and hinge reduced order modeling approach \cite{filipov2017bar} as implemented in MERLIN2 \cite{liu2018highly} was used to model the potential energy of the system. This model captures the in-plane normal and shear stiffness of the panel with a truss structure, and the bending and folding stiffness of origami with torsional springs along the bars (as demonstrated in Figure \ref{fig:MERLIN}). 

The total potential energy in the discretized form for a given loading condition will be a summation of strain energy of bar deformations ($U_{S}(x)$), strain energy at bending hinges ($U_{B}(x)$), strain energy at folding hinges ($U_{F}(x)$) minus the work done by the applied force ($f^{T}x$). More details regarding the derivation of potential energies, internal forces and stiffness can be found in \cite{liu2017nonlinear, liu2018highly}. Thus, total potential energy can be written as

    \begin{equation}
    P_{E} = U_{S}(x) + U_{B}(x) + U_{F}(x) - f^{T}x
    \label{eqn:Pe}
\end{equation}

where $x$ is the vector containing the displacements of the degrees of the freedoms (DOFs) and $f^{T}x$ represents the potential energy due to the applied force.

The kinetic energy of the structure will depend on the current position as well as the configuration of the panels, making its estimation and further computations very hard to analytically estimate. Therefore, in this study, the possibility of using distributed mass formulations is explored. The total kinetic energy can therefore be expressed as:

\begin{equation}
    K_E = \sum_{i=1}^{N} \sum_{j=1}^{N} 0.5m_{ij}v_{i}v_{j}
    \label{eqn:Ke}
\end{equation}

where $v_i$ is the velocity in the $i^{th}$ degree of freedom, $m_{ij}$ is the mass lumped between DOFs $i$ and $j$ and $N$ is the total number of DOFs in the system.  

In addition, the damping energy in this study is estimated assuming the nature of damping is linear proportional damping. Therefore, damping energy is expressed as:

\begin{equation}
    D_E = \sum_{i=1}^{N} \sum_{j=1}^{N} 0.5c_{ij}v_{i}v_{j}
    \label{eqn:De}
\end{equation}

where, $c_{ij}$ is the damping coefficient between DOFs $i$ and $j$. 

The Lagrange equation can be used to derive the equations of motion. The Lagrange equation states that for each DOFs $i$, 

\begin{equation}
  \pdv{L}{t}{\dot{x_i}} - \pdv{L}{x_i} + \pdv{D}{\dot{x_i}} = 0
  \label{eqn:Le}
\end{equation}

Combining the equations \ref{eqn:L}, \ref{eqn:Pe}, \ref{eqn:Ke} and \ref{eqn:De} in \ref{eqn:Le}, for the $i^{th}$ DOF, we get

\begin{align}
    \pdv{K_E}{t}{\dot{x_i}} + \pdv{P_E}{x_i} + \pdv{D}{\dot{x_i}} = 0 \\
    m_{ij} \Ddot{x_j} +  c_{ij} \dot{x_j} + F_{int,i} = f_{i}(t)
    \label{eqn:eom}
\end{align}

Equation \ref{eqn:eom} can be expressed in the matrix notation as:
\begin{equation}
    \bm{M} \bm{\Ddot{x}} + \bm{F_D(x, \dot{x})} + \bm{F_{int}(x)} = \bm{p_{ext}}
    \label{eqn:eom2}
\end{equation}
where, $\bm{F_D(x, \dot{x})} = \bm{C \dot{x}}$ for using the assumption of linear proportional damping and $\bm{F_{int}(x)}$ is computed using the bar and hinge model. 

The procedure followed to perform the dynamic simulations is discussed next. The geometric parameters such as width, length, panel angle and fold angle of origami was input to MERLIN. MERLIN then determines the location of the nodes and performs the requisite discretization. The support conditions and constitutive model parameters were then input to MERLIN, which along with discretization computes the stiffness matrix as a function of nodal displacements. Subsequently, the global mass matrix was assembled by combining the mass matrices of individual panels according to the given mass distribution. The global damping matrix was computed by combining the stiffness and mass matrices using the proportional damping formulation. Then, using the given loading time series and initial conditions across various nodes, time integration of Equation \ref{eqn:eom2} was performed using fourth-order Runge-Kutta method \cite{dormand1980family} to determine the structural response. Note that implicit methods can also be used to perform the time integration instead of fourth-order Runge-Kutta method.

\section{Different mass distribution formulations}
\begin{figure}
 \centering
  \includegraphics[width=0.8\linewidth]{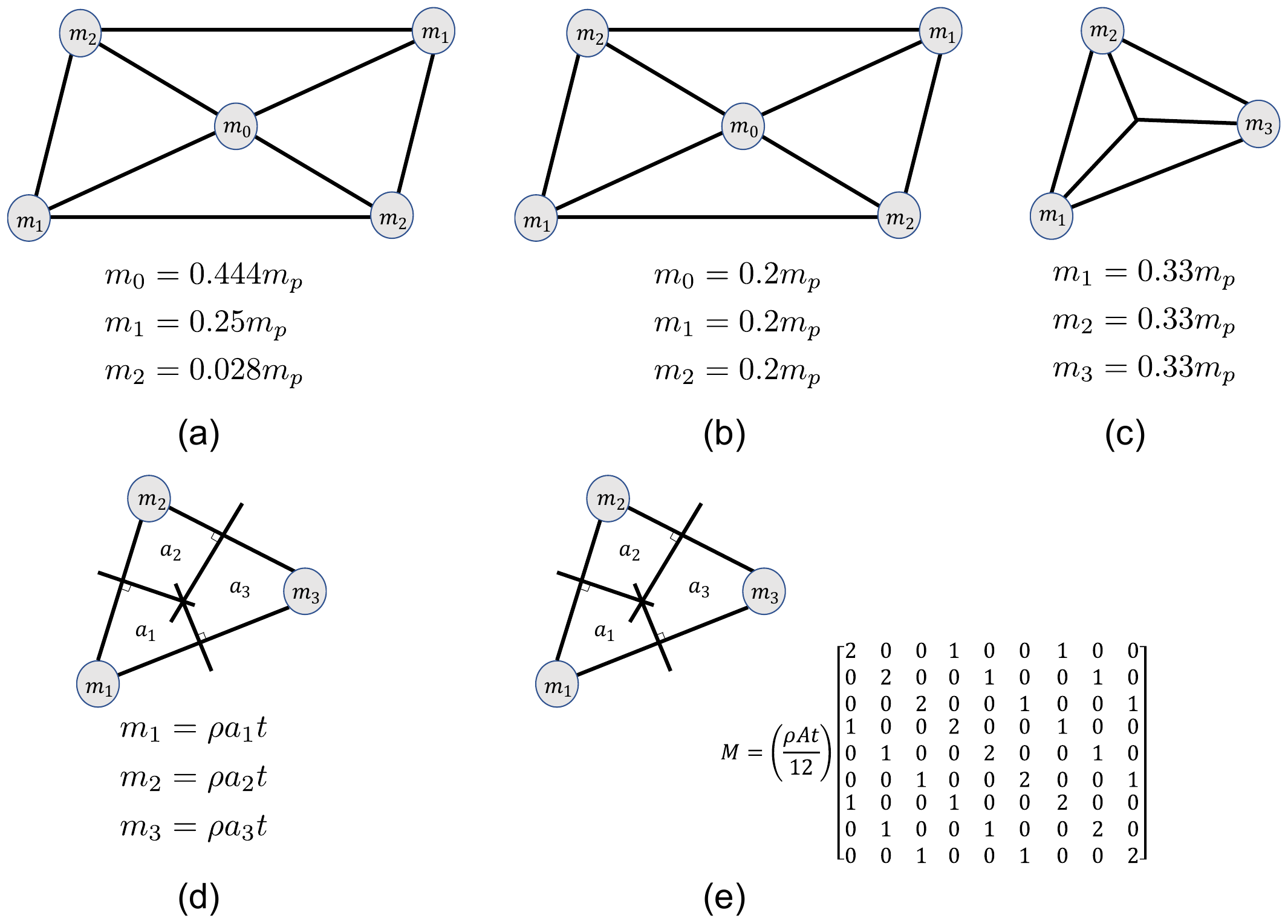}
  \caption{Mass lumping techniques: (a) Equi-inertial mass matrix (proposed by Xia and Wang \cite{xia2019dynamics}), (b) Panel mass distribution, (c) Triangle centroid mass distribution, (d) Triangle circumcenter mass distribution, and (e) Consistent triangle mass matrix.}
  \label{fig:mass_mat}
\end{figure}

In this study, we use a mass distribution model to distribute the mass to different DOFs within a panel. Various mass distribution models, based on varied reasoning, can be potentially used for the distribution. In this section, different mass distribution techniques along with the corresponding rationale and advantages are described. The different techniques considered are: (a) Mass discretization proposed by Xia and Wang \cite{xia2019dynamics}, (b) Panel mass distribution, (c) Triangular centroidal mass distribution, (d) Triangle circumcenter mass distribution and (e) Consistent triangle mass matrix used in Finite element literature.
Figure \ref{fig:mass_mat} demonstrates the mass distribution in each of these techniques.

\subsection{Equi-inertial mass matrix}
Xia and Wang \cite{xia2019dynamics} propose a mass distribution technique for a panel by balancing the moment of inertia of the panel in its principle directions with the moment of inertia of point mass assemblages located at the nodes (shown in Figure \ref{fig:mass_mat}). They find that when the mass is distributed as $[0.444, 0.25, 0.028]$ of the mass of the panel at $m_0, \ m_1 \ and \ m_2$ locations, the error between the moment of inertias are $[0.8\%, 0\%, 0.2\%]$. This mass matrix formulation is therefore referred to as the `Equi-Inertial mass matrix' for the rest of the discussion.  

\subsection{Panel mass distribution}
Panel mass distribution represents another approach for distributing the mass within a panel. In this technique, an equal amount of the panel mass is allocated to all the nodes (as shown in Figure \ref{fig:mass_mat}). In this approach, the geometry of the panel is not accounted for and therefore, various panel geometries lead to the same mass distribution as long as the mass of the panel is the same. This is the primary disadvantage of the distribution. \\

Since there are some origami structures that already have a triangular faceted geometry such as the Kresling pattern, the above distribution strategies might not be directly applicable for such cases. On the hand, every panel is further faceted to triangles in bar and hinge reduced order modeling. Therefore, it might be beneficial to consider triangular mass distribution techniques so that the mass of a triangular facet can be distributed to its corresponding nodes and then combined to form the mass matrix for the entire panel. With this consideration, three distribution techniques based on triangles were considered that are described next.

\subsection{Triangle centroid mass distribution}
Triangle centroid mass distribution is a naive approach to distribute the mass about the centroid of the triangle. This will result in an equal mass distribution to all the nodes of the triangle. In accordance with this mass distribution, different triangular shapes with the same mass will lead to the same mass distribution which is the major drawback of this approach. 

\subsection{Triangle circumcenter mass distribution}
The triangle circumcenter mass formulation was selected to account for the effect of geometry on mass distribution. In this approach, the perpendicular bisectors of the three sides are constructed which intersect at the circumcenter and divide the triangle into three quadrilaterals (see Figure \ref{fig:mass_mat}). The area enclosed by the quadrilateral containing a node was considered to be its tributary area. These areas were then used to allocate the mass as $m_i = \rho a_i t$, where $\rho$ is the density of the material, $a_i$ is the tributary area of $i^{th}$ node, and $t$ is the thickness of the panel.

\subsection{Consistent triangle mass matrix}
Triangle consistent mass matrix is often used in finite element modeling literature to capture the mass distribution of a triangular element. In this formulation, it is assumed that the displacements ($u(X)$) and velocities ($\dot{u}(X)$) in the triangular element vary linearly within a triangular element with the same interpolations functions as 
\begin{align}
    & \bm{u(X)} = \bm{N^t U} \\
    & \bm{\dot{u}(X)} = \bm{N^t \dot{U}}
\end{align}
where, $\bm{N}$ is the matrix containing the interpolation functions, $\bm{U}$ is the nodal displacement vector, $\bm{\dot{U}}$ is the nodal velocity vector and $\bm{X}$ is any point within the triangle. Thus, the kinetic energy of the element can be written as:
\begin{equation}
    K_{E} = \frac{1}{2} \int \int \int_{V^{e}} \rho \dot{\bm{u}}^{T} \dot{\bm{u}} dV = \frac{1}{2} \dot{\bm{U}}^{T} \left( \int \int \int_{V^{e}} \rho \bm{N}^t \bm{N} dV \right) \dot{\bm{U}}
\end{equation}
where, $\rho$ is the density of the material of triangle and $V^{e}$ is the volume of the element.

\begin{equation}
    M = \frac{\partial^{2} K_{E}}{\partial \dot{\bm{U}} \partial \dot{\bm{U}}} = \int \int \int_{V^{e}} \rho \bm{N}^t \bm{N} dV
\end{equation}
On performing the integration, the resultant consistent mass matrix for a triangular element is given by
\begin{equation}
    M = \frac{\rho A t}{12} \begin{bmatrix}
          2 & 0 & 0 & 1 & 0 & 0 & 1 & 0 & 0 \\
          0 & 2 & 0 & 0 & 1 & 0 & 0 & 1 & 0 \\
          0 & 0 & 2 & 0 & 0 & 1 & 0 & 0 & 1 \\
          1 & 0 & 0 & 2 & 0 & 0 & 1 & 0 & 0 \\
          0 & 1 & 0 & 0 & 2 & 0 & 0 & 1 & 0 \\
          0 & 0 & 1 & 0 & 0 & 2 & 0 & 0 & 1 \\
          1 & 0 & 0 & 1 & 0 & 0 & 2 & 0 & 0 \\
          0 & 1 & 0 & 0 & 1 & 0 & 0 & 2 & 0 \\
          0 & 0 & 1 & 0 & 0 & 1 & 0 & 0 & 2 \\
        \end{bmatrix}
\end{equation}
A detailed derivation of the triangle consistent mass distribution can be found in An introduction to the finite element method by J.N.Reddy \cite{reddy1993introduction}. Note that this mass matrix is also invariant to the triangle geometry unlike the Triangle circumcenter mass distribution.

\section{Numerical simulations and results}
The previous section described various mass distribution techniques that were selected for potential implementation into the dynamic modeling framework. To gauge their effectiveness as well as the effectiveness of the dynamic modeling approach itself in capturing the dynamics, a benchmark model is needed. For this, finite element modeling is used and the results from the proposed dynamic model with different mass matrix formulations are compared with the results from finite element modeling. Four different scenarios consist of various loading and boundary conditions, geometric properties and deformation modes. This section first describes the procedure for finite element modeling and then presents the results of different scenarios considered.

\subsection{Finite element modeling}
In this study, finite element modeling was used as a benchmark to gauge various mass distribution techniques. For this, the ABAQUS/Standard was used for performing the finite element simulations. A procedure described in the literature \cite{xu2021torsional, grey2019strain, grey2020mechanics} was followed for the modeling of origami structures. Under this approach, the panel is modeled using linear and quadratic shell elements (i.e. S4R and S3 shell elements in ABAQUS) and the connections between the panels are modeled using hinge connector (i.e. CONN3D2 connector elements in Abaqus). The isotropic elastic material model is used for the behavior of the panel material. 

MERLIN calibrates the bending stiffness ($k_{B}$), bar stiffness ($k_{S}$) and fold stiffness ($k_{F}$) using Young's modulus, Poisson's ratio and L-scale factor ($L*$). The L-scale factor is defined as the ratio of the bending modulus of the sheet multiplied by the length of the fold and the rotational stiffness of the fold line \cite{filipov2017bar} (i.e. $L*=Lf*k/K_{f}$, where $k$ is the bending modulus and $K_{f}$ is the fold stiffness). To be able to compare to the proposed dynamic formulation, the same Young's modulus, and Poisson ratio were used for both models. The stiffness of hinges in finite element model was obtained by dividing the fold stiffness of MERLIN ($k_f$) with the number of connector nodes in the FEM ($k_{conn} = k_f/N_{nodes}$, where $N_{nodes}$ is the number of nodes connected between the panels as the model is prepared such that a hinge is present at each of the edge node) which ensured that the fold stiffness in both models remained the same. In addition, the loading and the boundary conditions in the models were also the same. The mesh density was chosen after a mesh sensitivity analysis and the size was chosen as 0.1 since the change was less than 0.5\% after increasing the mesh density by 50\%. 

\subsection{Results}
This section presents the results of four test cases that were tried. The cases include: (a) 
a simple fold, (b) Miura-ori with X-directional loading, (c) Miura-ori with Y-directional loading, and (d) Miura-ori with a high L-scale factor ratio leading to a fold-dominated response. These four cases represent different geometries, loading and boundary conditions, and material properties and thus allow us to gauge the consistency of performance of different mass distributions. For each of case study, quasi-static simulations were first performed to ensure that the stiffness between MERLIN and FEM agreed with each other. This was then followed by a dynamic loading scenario with an impulse type loading, which ensured that system properties (namely, mass and stiffness) governed the response.   

\begin{figure}
 \centering
  \includegraphics[width=0.8\linewidth]{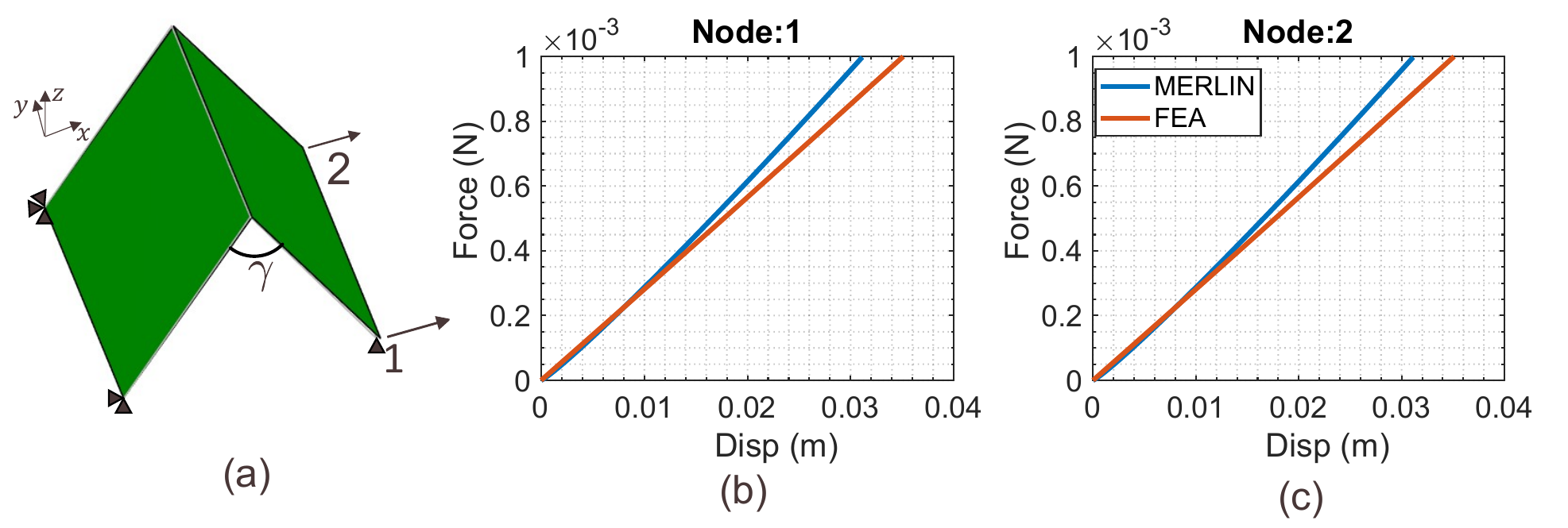}
  \caption{(a) The setup of simulation for simple fold, showing geometry, loading and boundary conditions. (b) Force displacement of Node 1. (c) Force displacement of Node 2.}
  \label{fig:qs_simple_fold}
\end{figure}

\begin{figure}[t]
 \centering
  \includegraphics[width=0.5\linewidth]{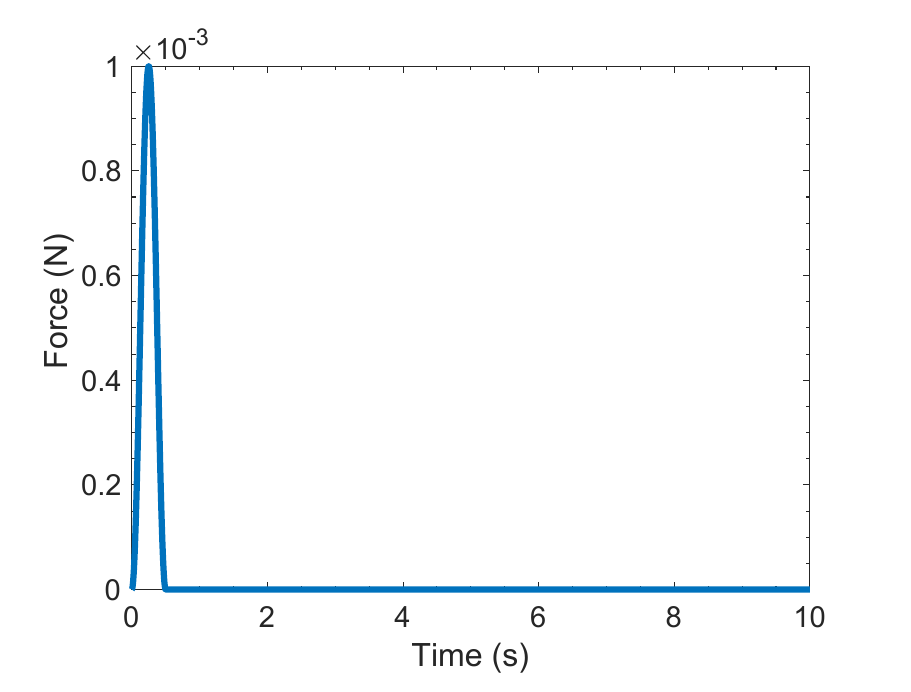}
  \caption{The loading applied for the dynamic simulation of a simple fold at nodes 1 and 2.}
  \label{fig:case_a_load}
\end{figure}

\begin{figure}
 \centering
  \includegraphics[width=0.95\linewidth]{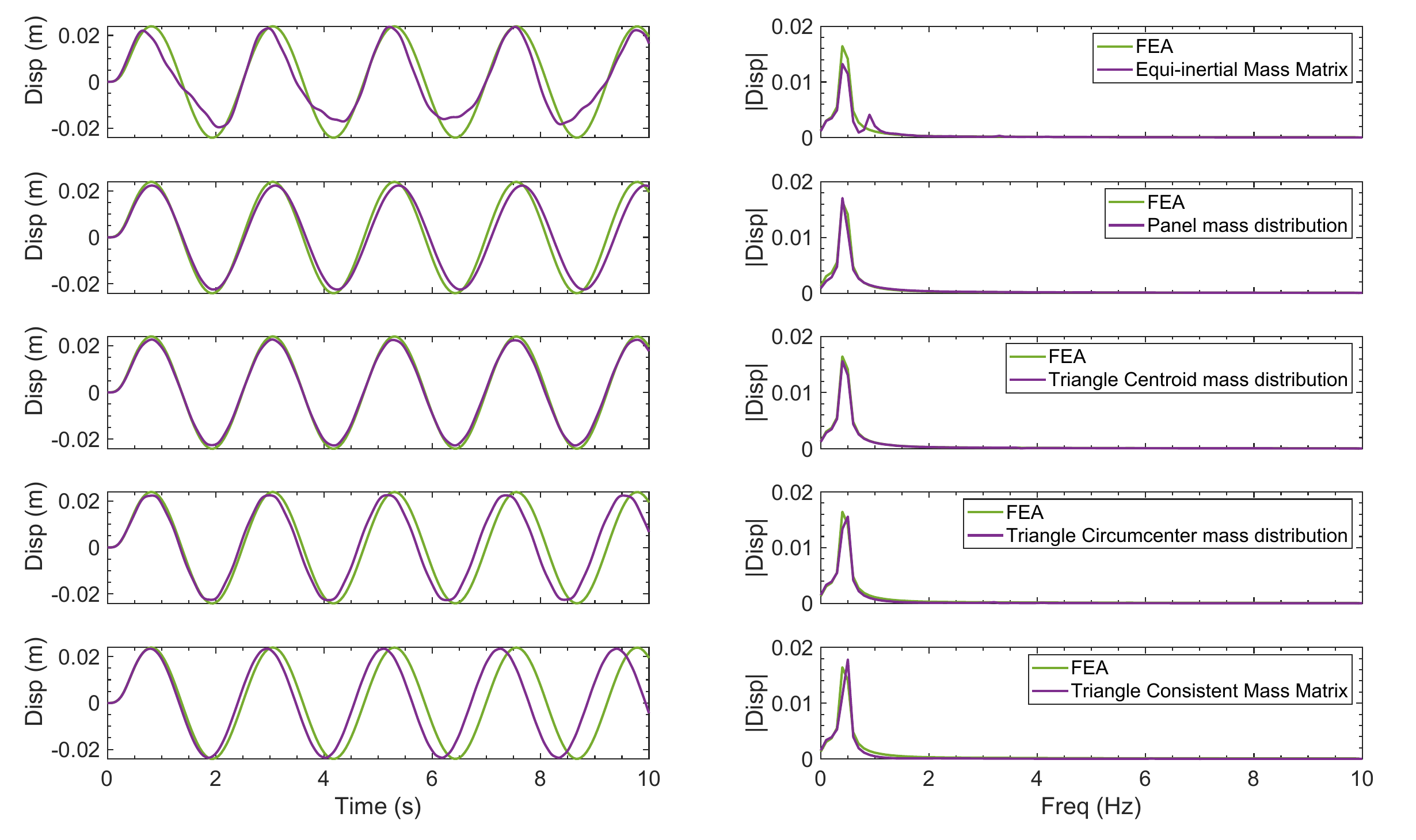}
  \caption{The response of Node 1 under different mass distribution formulations: Left - Displacement response and Right - corresponding frequency response.}
  \label{fig:case_a_node_1}
\end{figure}

\subsubsection{Case study 1: Simple fold}
The first case study was the simulation of a simple fold. The dimensions of the origami square panel are $1 \ m$ by $1 \ m$ with a fold angle of $60 \degree$ with a thickness of $0.01 \ m$. The geometry, loading and boundary conditions are shown in Figure \ref{fig:qs_simple_fold}. Young's modulus of the material was $10^6 \ N/m^2$ and Poisson's ratio was $0.25$. The ratio of L-scale factor was taken to be $1$. These values were chosen in line with the study by Filipov et al \cite{filipov2015origami}. Using the calibration procedure of MERLIN \cite{liu2018highly}, the folding stiffness is $0.0639 \ Nm/rad$. To compare the influence of various mass distribution techniques, we must first ensure that the stiffness between the two models agree with each other. Therefore, we first perform a quasi-static analysis. A concentrated force of magnitude $10^{-3}$ was applied on nodes 1 and 2. The results of the static analysis are shown in Figure \ref{fig:qs_simple_fold}(b) and (c). We can see that there is a good agreement between the stiffness of FEA and MERLIN.

\begin{figure}
 \centering
  \includegraphics[width=0.95\linewidth]{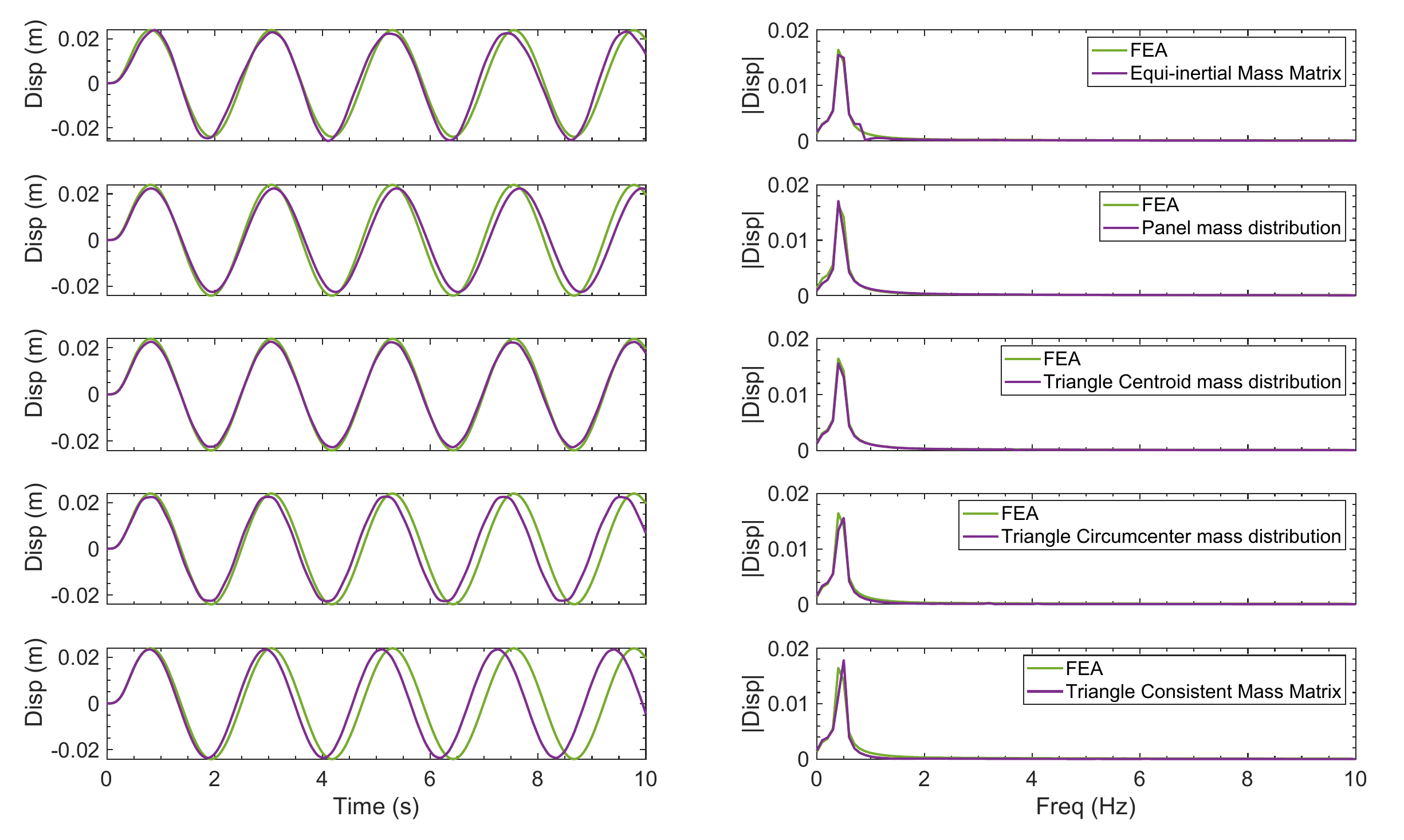}
  \caption{The response of Node 2 under different mass distribution formulations: Left - Displacement response and Right - corresponding frequency response.}
  \label{fig:case_a_node_2}
\end{figure}

Next, the dynamic behavior was studied. For this, the simple fold was subjected to the loading shown in Figure \ref{fig:case_a_load}. The loading consists of an impulse followed by no loading for the rest of the time. This load was applied on node 1 and node 2, while the nodes on the other end were kept fixed (i.e. loading and boundary were the same as that of the quasi-static loading scenario). The response of the system to such a loading after the removal of force will be governed by its impulse response behavior, which depends only on system properties. Comparing the impulse response characteristics is a direct way to compare the system properties as the two are closely related. The results from this load application are shown in Figure \ref{fig:case_a_node_1} and \ref{fig:case_a_node_2}, which show the displacement of Nodes 1 and 2 using different mass matrix formulations. This figure also shows the frequency response of these responses. We can see that most of the mass matrices are able to capture the dynamics accurately. The frequency content of systems containing different mass matrices indicates that most of the mass matrices were able to capture the frequency very well for both nodes. Note that even a small difference in frequency estimation might seem like a deviation in the time domain as there is a very small amount of damping and therefore, we use the relative difference in the frequency content when quantifying the error between the various formulations and FEA. Specifically, the error metric is defined as the norm of the difference in frequency content divided by the norm of FEA time series frequency content for the nodes. The mean error metric is the mean of the error metrics of individual nodes. The mean error metric for the five mass matrix formulations is shown in Table \ref{table:table1} for the four case studies. Triangle centroid mass distribution was the best-performing mass matrix in this case, followed by Panel and Triangle circumcenter mass distributions which were just slightly off in their frequency content. Another interesting observation was that the responses of nodes 1 and 2 were different indicating asymmetry in the system with an Equi-inertial mass matrix. This is because of the asymmetrical nature of the mass matrix itself. This asymmetry resulted in a second peak frequency response not seen in other mass models (see Figure \ref{fig:case_a_node_1}).

\begin{table}[t]
\caption{Mean error metric for different mass distributions}
\label{table:table1}
\begin{tabular}{|l|lllll|}
\hline
\multirow{2}{*}{Case} & \multicolumn{5}{c|}{Mass distribution}                                                                                                                                       \\ \cline{2-6} 
                      & \multicolumn{1}{l|}{Equi-inertial} & \multicolumn{1}{l|}{Panel}  & \multicolumn{1}{l|}{Triangle centroid} & \multicolumn{1}{l|}{Triangle circumcenter} & Triangle consistent \\ \hline
1                     & \multicolumn{1}{l|}{0.1970}        & \multicolumn{1}{l|}{0.1556} & \multicolumn{1}{l|}{0.0620}            & \multicolumn{1}{l|}{0.1698}                & 0.2931              \\ \hline
2                     & \multicolumn{1}{l|}{0.6012}        & \multicolumn{1}{l|}{1.1112} & \multicolumn{1}{l|}{0.7543}            & \multicolumn{1}{l|}{0.5347}                & 0.3966              \\ \hline
3                     & \multicolumn{1}{l|}{1.2393}        & \multicolumn{1}{l|}{0.7285} & \multicolumn{1}{l|}{0.3764}            & \multicolumn{1}{l|}{0.2257}                & 0.9373              \\ \hline
4                     & \multicolumn{1}{l|}{0.3445}        & \multicolumn{1}{l|}{0.5290} & \multicolumn{1}{l|}{0.4195}            & \multicolumn{1}{l|}{0.1957}                & 0.2509              \\ \hline
\end{tabular}
\end{table}


In the interest of brevity, case studies 2 and 3 have been presented in the supplemental section. In the second case study, a more complicated geometry was chosen to test the consistency of these formulations across different geometries. With this consideration, Miura-ori unit cell was chosen as it is one of the most widely used origami geometry for engineering applications. The panel dimensions of Miura-ori were taken as $1 \ m$ by $1 \ m$ with a thickness of $0.01 \ m$. The dihedral fold angle, $\theta$ was retained at $60 \degree$ and the facet angle $\gamma$ was also taken as $60 \degree$. The Young's modulus and Poisson's ratio were retained as $10^6 N/m^2$ and $0.25$ respectively, while the L-scale factor was chosen as $1 \ m$. The mean error metrics for the second case study is shown in Table \ref{table:table1}. The Triangle consistent mass distribution was the best performing mass distribution followed by Triangle circumcenter mass distribution. The third case we consider is the loading of the Miura-ori unit cell in a different loading direction. This further reinforces the consistency of different mass formulations for various displacement modes in addition to various geometries as well. The mean error metrics for the third case study are shown in Table \ref{table:table1}. The best performing mass distribution in this case was Triangle circumcenter mass distribution followed by Triangle centroid mass distribution. The time series results and other details can be found in the supplemental section.

\subsection{Case Study 4: Influence of L-scale factor in Miura-ori X directional loading}

\begin{figure}
 \centering
  \includegraphics[width=0.65\linewidth]{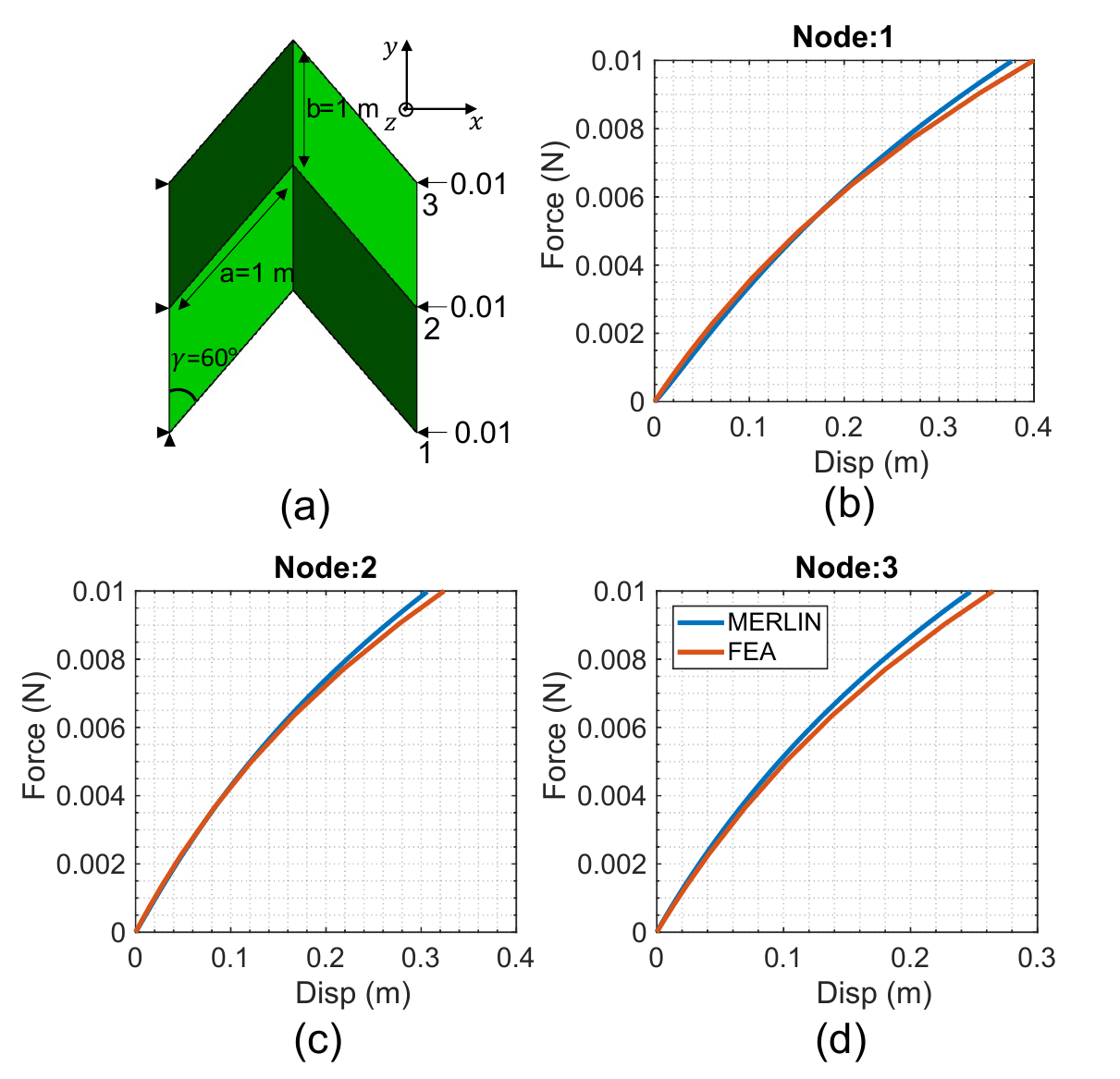}
  \caption{Quasi-static simulation of Miura-Ori with high L-scale factor under X-direction loading: (a) Force displacement of node 1, (b) Force displacement of node 2 and (d) Force-displacement of node 3.}
  \label{fig:case_d_qs}
\end{figure}

The last case study was performed to check the influence of the L-scale factor on the choice of mass matrix formulation. The L-scale factor ($L^{*}$) dictates the relative stiffness between the bending and the folding deformation of panels, with a low L-scale factor ($L^{*} \leq 1$) leading to bending-dominated response and a high L-scale factor ($L^{*} > 4$) leading to a folding-dominated response. By varying this parameter, we can ensure that our inferences on various mass formulations remain consistent across different dominant modes of deformation (which was different in this case study than in Case Study 2). For this case study, all the parameters, loading, and boundary conditions were kept the same as in case study 2, except for the L-scale factor, which was increased to 5. Since the deformations are folding-dominated, under the application of the same amount of force, the deformations are significantly higher than in Case Study 2 (almost a three-fold increase). Further, we can also see some amount of nonlinearity in the stiffness. We first again compare the quasi-static behavior of the MERLIN model with FEM, shown in Figure \ref{fig:case_d_qs}. The agreement between the two methods is much higher than in the other case studies. The dynamic simulations were then performed to gauge these different mass matrix formulations, which were subjected to much higher deformations and non-linearity in terms of stiffness.

\begin{figure}
 \centering
  \includegraphics[width=0.95\linewidth]{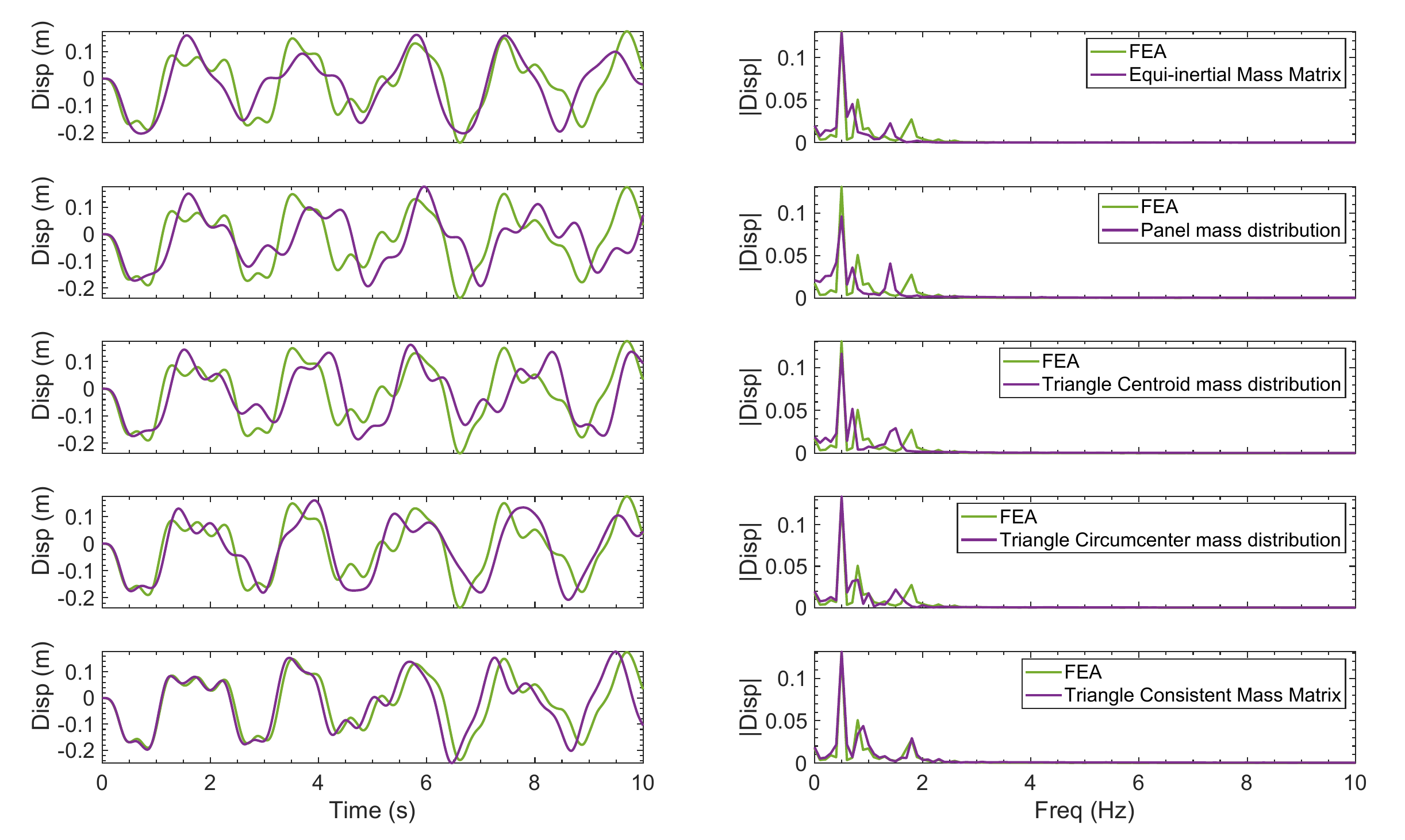}
  \caption{Results from dynamic simulation under X-direction loading of Miura Ori with high L-scale factor: Left - Time series of displacement of Node 1 with different mass matrices and Right - Corresponding frequency content.}
  \label{fig:case_d_qs1}
\end{figure}

\begin{figure}
 \centering
  \includegraphics[width=0.95\linewidth]{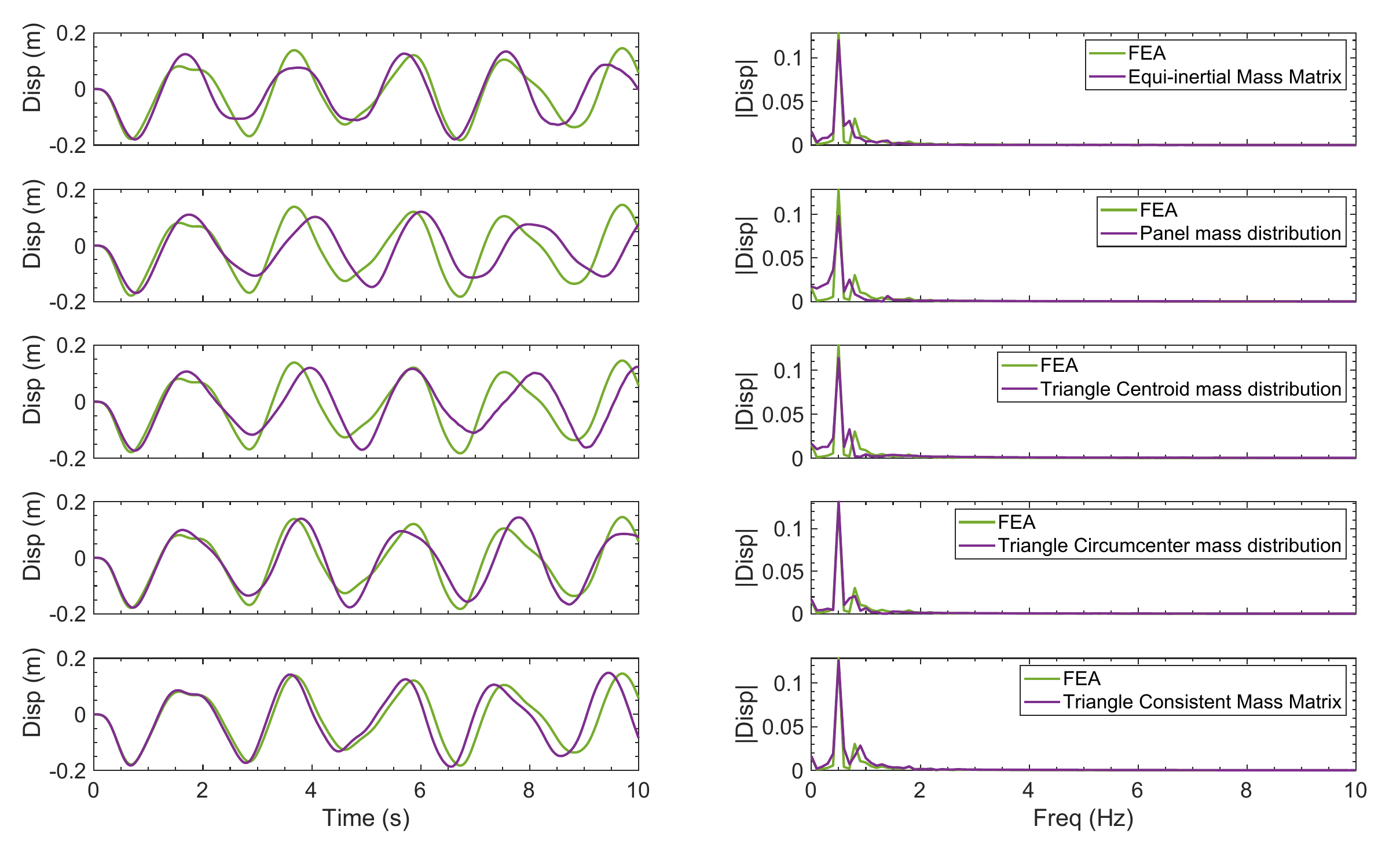}
  \caption{Results from dynamic simulation under X-direction loading of Miura Ori with high L-scale factor: Left - Time series of displacement of Node 2 with different mass matrices and Right - Corresponding frequency content.}
  \label{fig:case_d_qs2}
\end{figure}

\begin{figure}
 \centering
  \includegraphics[width=0.95\linewidth]{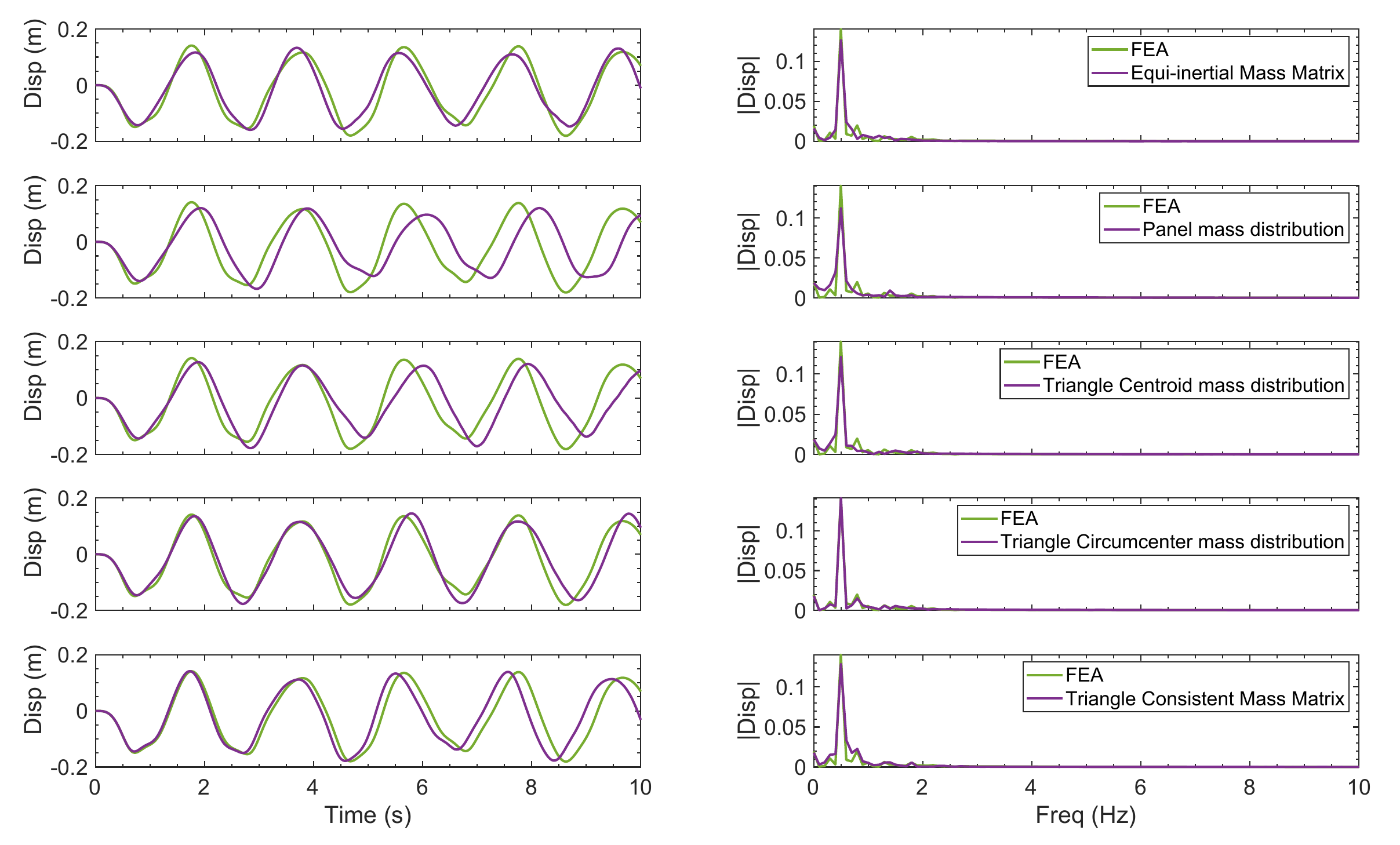}
  \caption{Results from dynamic simulation under X-direction loading of Miura Ori with high L-scale factor: Left - Time series of displacement of Node 3 with different mass matrices and Right - Corresponding frequency content.}
  \label{fig:case_d_qs3}
\end{figure}

The results from the dynamic simulation for different mass matrices and FEM are shown in Figures \ref{fig:case_d_qs1}, \ref{fig:case_d_qs2} and \ref{fig:case_d_qs3}. In the response on node 1, the Triangle consistent mass matrix was able to capture all the dynamics with a slight difference, owing to a slight estimation error in the second frequency peak (around 1 Hz). The other mass matrices except the Panel mass distribution were able to capture the low-frequency behavior of the dynamics well, while they had some error in estimating the higher frequencies. However, for nodes 2 and 3, Triangle circumcenter distribution was able to capture most of the dynamics, with a slight error in the estimation of the second frequency peak. The mean error metrics for the fourth case study are shown in Table \ref{table:table1}. Triangle centroid distribution was the second best with a similar trend. However, for these nodes, the other formulations were not able to estimate the peak frequency accurately.  

Overall, from the four case studies it was observed that Triangle circumcenter distribution was able to capture most of the dynamics consistently and can be a good candidate for use. It was able to capture most of the low-frequency content very well across all the four cases investigated while the other mass formulations were not able to capture the peak frequency accurately in at least one of the cases. Moreover, for most of the cases, the error between model with Triangle Circumcenter Mass formulation and FEA resulted from a slight inaccuracy in the estimation of higher frequency. These differences might further reduce with the incorporation of appropriate damping.

\section{Experimental validation}
The previous sections described the dynamic formulation and different mass distribution strategies that were chosen, and the numerical investigation of these strategies was presented. From the numerical investigation, it was seen that Triangle circumcenter mass distribution was able to capture most of the dynamics for various loading and boundary conditions, geometric as well as material properties cases. In this section, experimental validation of the dynamic formulation is presented along with the details of the specimen tested and the methodology for quasi-static and dynamic testing. 

\subsection{Specimen details} \label{sec:mo_det}
\begin{figure}
 \centering
  \includegraphics[width=0.6\linewidth]{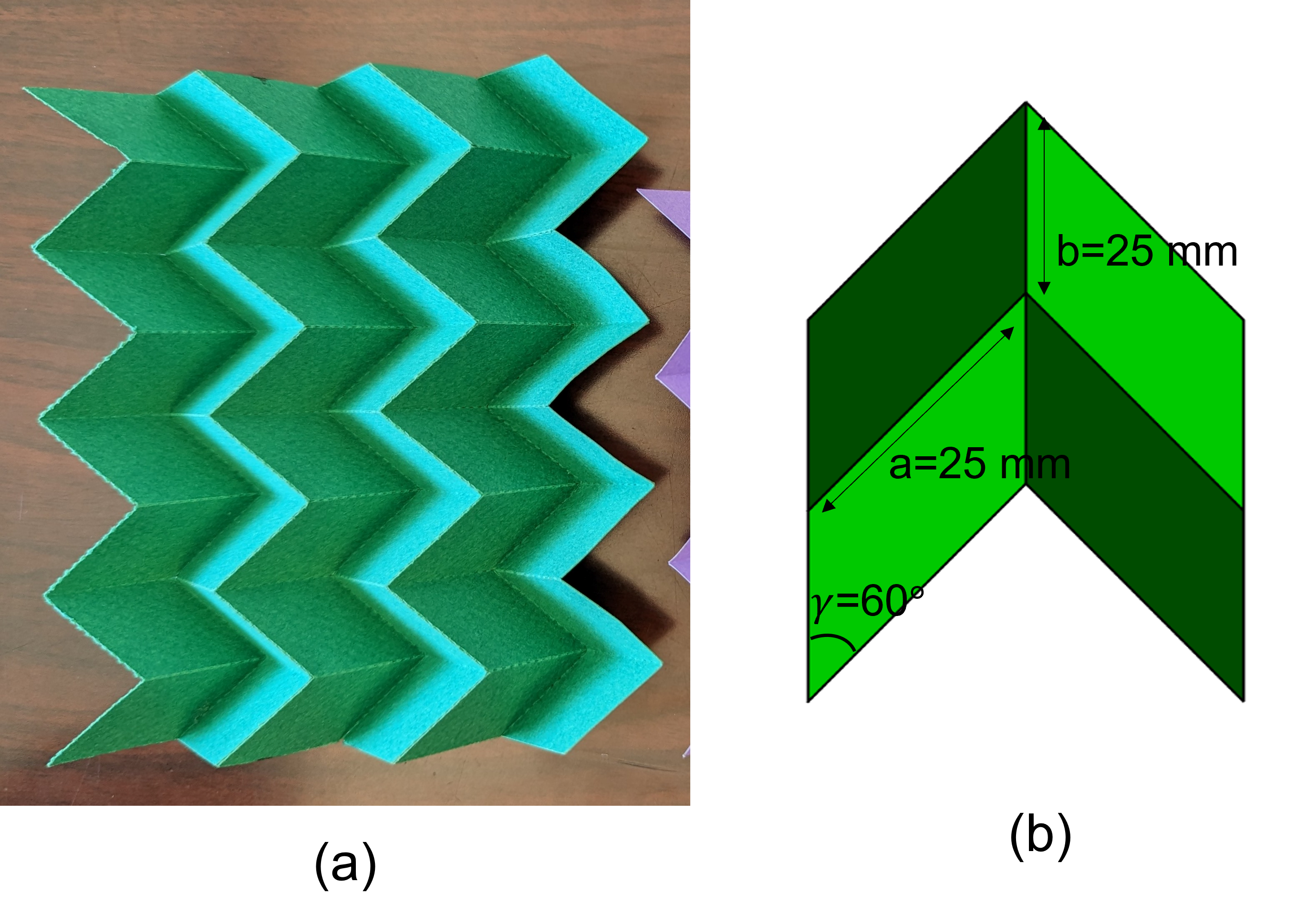}
  \caption{(a) Manufactured origami sample. (b) Geometric dimensions of Miura-ori unit cell.}
  \label{fig:mo_sample}
\end{figure}

A Miura-ori sample was used in this study for experimental validation of the dynamic model. The sample was manufactured using Canson Mi-Teintes craft paper with a thickness of $0.24 \ mm$ and Young's modulus of $1219 \ MPa$ \cite{liu2020big}. For the manufacture of the specimen, crease patterns were formed on the paper by adding perforations using an electronic cutting machine, which then allowed the paper to be folded to the shape of the origami. Further details of the manufacture of the specimen can be found in Liu et al \cite{liu2020big}. The photograph of the sample is shown in Figure \ref{fig:mo_sample}(a), while the geometric dimensions of the unit cells are shown in Figure \ref{fig:mo_sample}(b). As seen from the image, the specimen had 3 cells in the horizontal direction and 4 cells in the vertical direction. In this study, the full-field dynamic response of the specimen in the experiment was compared to the prediction of the analytical model. To obtain the full-field response, the video of the experimental response of the specimen was recorded using a camera and then computer vision-based KLT feature point tracking \cite{lucas1981iterative, tomasi1991detection} with Harris corner points \cite{harris1988combined} was used to obtain the response of the specimen. The front-facing nodes of the specimen were the data acquisition nodes and were used for all the comparisons. To facilitate easy detection and trackability of these corner points, small paper sheets were attached to each of the front-facing nodes of the specimen.  

\subsection{Quasi-Static experiments}
After the manufacture and preparation of specimens, quasi-statics tests were first performed. The quasi-static tests were performed with two objectives: (1) calibrate the folding and bending stiffness of the hinges in the MERLIN model, and (2) verify that the MERLIN model is able to capture the stiffness of the specimen as was done in the empirical studies. 

\subsubsection{Experimental Setup}
\begin{figure}
 \centering
  \includegraphics[width=0.6\linewidth]{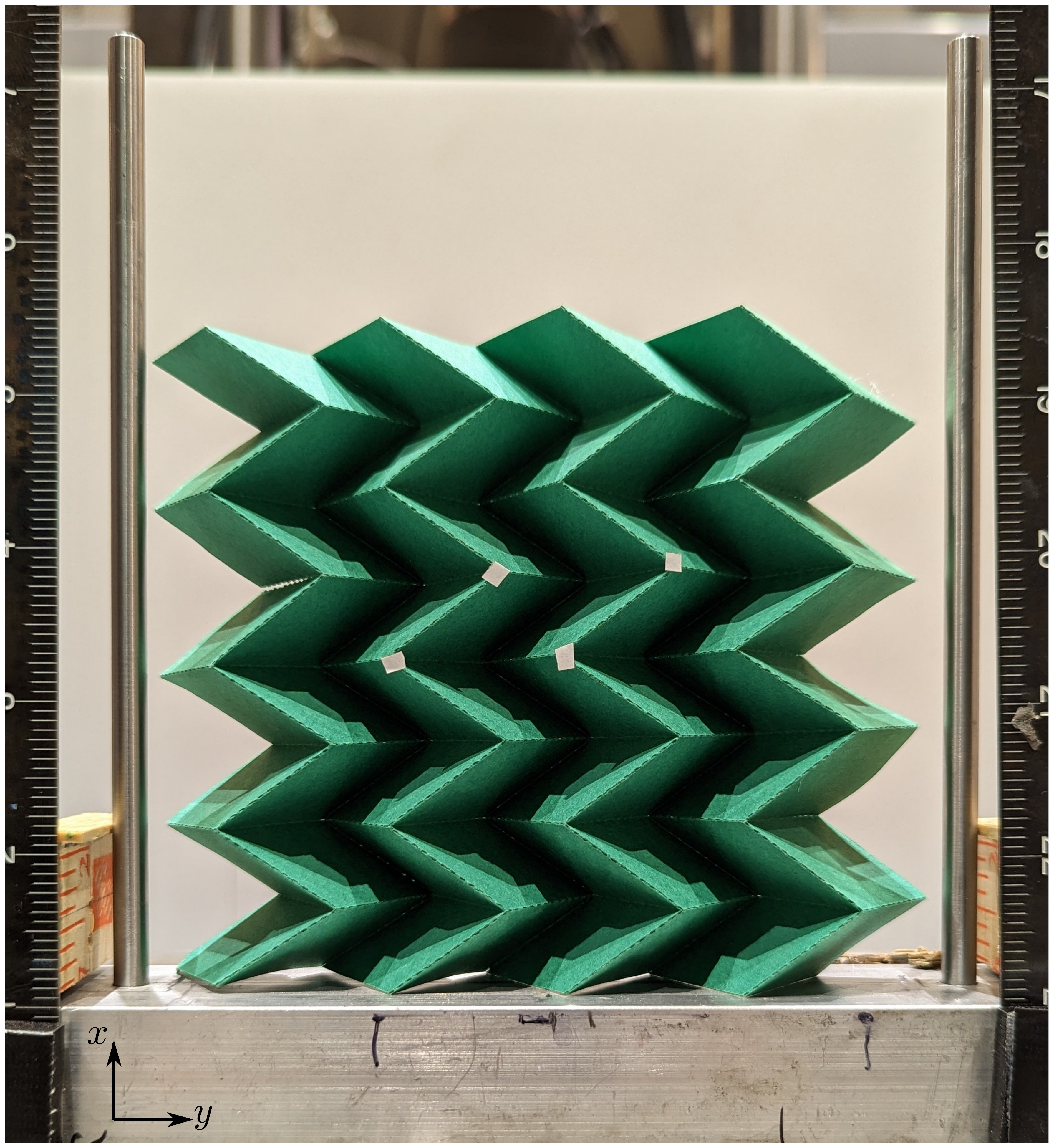}
  \caption{Experimental setup used for testing the origami samples.}
  \label{fig:orig_exp_setup}
\end{figure}

A loading apparatus was fabricated to perform the quasi-static testing of the specimen, which is shown in Figure \ref{fig:orig_exp_setup}. The setup was fabricated by installing two vertical rods on the flat aluminum base. The specimen was mounted on the aluminum base as shown in Figure \ref{fig:orig_exp_setup}. The base was regularly cleaned with alcohol to ensure that the friction between the specimen and the base remained low during the experiments. The two vertical rods were used to insert and load requisite weights onto the specimen. Two scales were installed for calibration from pixel units to real-world dimensions. A mobile camera was placed in front of the loading apparatus to acquire the video data (with a resolution of $2268 \times 4032$ pixels), which was then used to compute the displacements in the specimen. The camera was placed by ensuring that the plane of the camera was as parallel as possible to the plane of the Miura-ori specimen. 

\subsubsection{Experimental data and model calibration}

\begin{figure}
 \centering
  \includegraphics[width=0.7\linewidth]{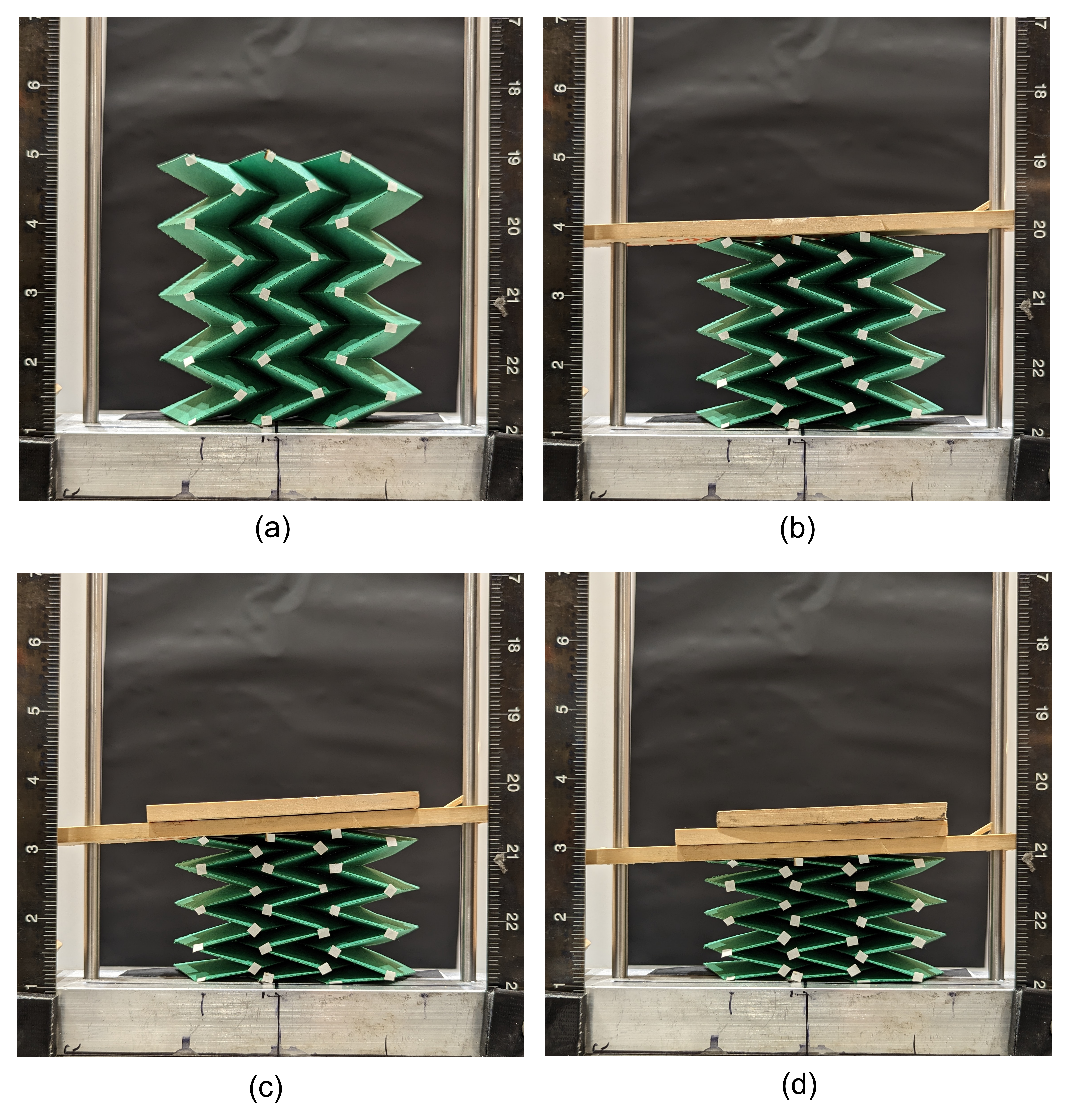}
  \caption{The displacement profiles of Miura-ori specimen for loading of: (a) 0 grams, (b) 12.75 grams, (c) 23.05 grams and (d) 31.85 grams.}
  \label{fig:orig_qs_data}
\end{figure}

The Miura-ori specimen described in section \ref{sec:mo_det} was incrementally loaded on the quasi-static loading apparatus using three loads of 12.75 grams, 10.3 grams, and 8.8 grams. The resultant deformation profiles of the origami sample are shown in Figure \ref{fig:orig_qs_data}. The corners from the deformation profiles were determined using the Harris corner detector to compute the experimental deformations across the sample. These deformations were then used to calibrate the material properties in MERLIN as well as to verify that MERLIN is able to capture the stiffness of the origami specimen.  

\begin{figure}
 \centering
  \includegraphics[width=\linewidth]{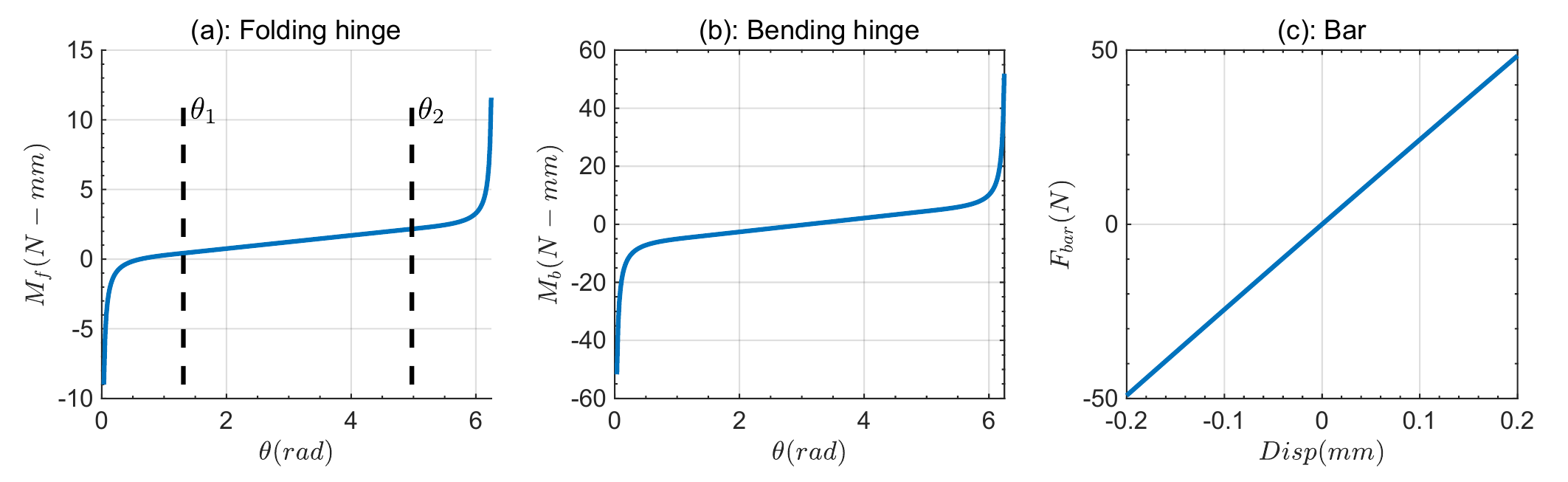}
  \caption{Constitutive models of (a) folding hinge $2-9$, (b) bending hinge $8-64$ and (c) bar $2-9$.}
  \label{fig:constit_relat}
\end{figure}

MERLIN model requires various geometric and material parameters as the input. The geometric parameters needed for the MERLIN model are $a = 25 \ mm, \ b = 25 \ mm, t = 0.24 \ mm$ and $\gamma =  60^{\circ}$, which are shown in Figure \ref{fig:mo_sample}. The input material properties that are needed for the MERLIN model are: Young's modulus ($E$), L-scale factor ($L^{*}$) (which represents the relative stiffness of folding hinges) and the stiffening angles of folding hinge, $\theta_{1}$ and $\theta_{2}$. Using the acquired experimental data, MERLIN parameters were calibrated and material properties were computed. The Young's modulus was determined by Liu et al\cite{liu2020big} for Canson Mi-Tientes paper as $1219 \ MPa$ and Poisson's ratio of $0.2$ was used in this study. From the calibration, it was found that ratio of the length scale factor ($L^{*}/L_{F}$) was 2.7. The folding hinge stiffening angles were found to be $\theta_{1} = 75^{o} (1.309 \ rad)$ and $\theta_{2} = 285^{o} (4.974 \ rad)$. The sample constitutive model for bars and hinges along with folding hinge stiffening angles (shown using dashed lines) are shown in Figure \ref{fig:constit_relat}. 

\begin{figure}
 \centering
  \includegraphics[width=0.7\linewidth]{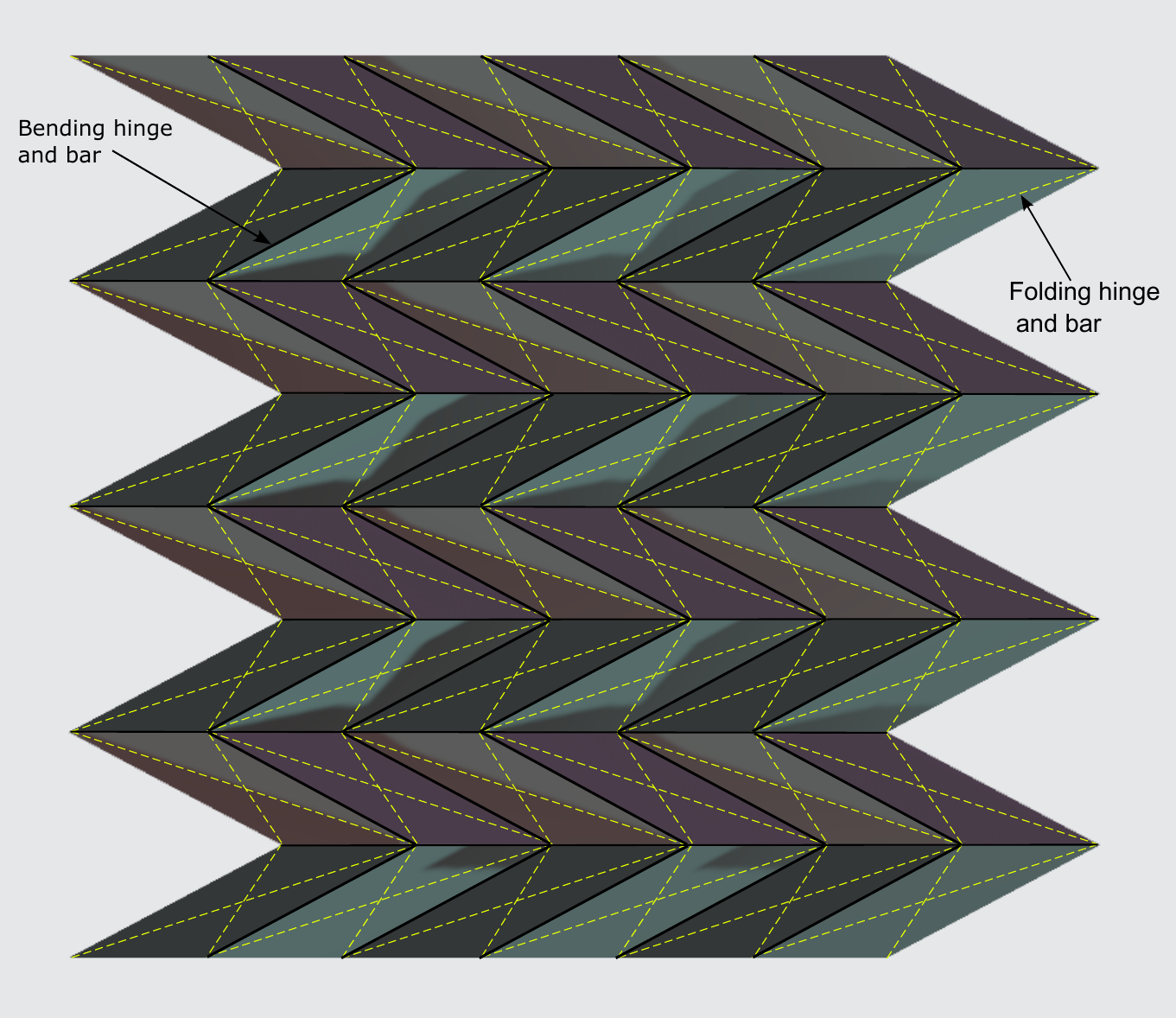}
  \caption{Discretization of Miura-ori specimen into bars, folding and bending hinges. The bars and folding hinges are shown using black lines while the bars and bending hinges are shown using yellow dashed lines.}
  \label{fig:miura_discrete}
\end{figure}

\subsubsection{Quasi-static results comparison}
The results from numerical simulations of quasi-static testing are compared to the experimental results in this section. The images taken during the experiments (shown in Figure \ref{fig:orig_qs_data}) were cropped and the location of corners was determined to compute the deformations in the specimen on the edge where the load is applied. Using the deformation profile of the specimen edge, quasi-static simulation was performed with displacement loading. Using the simulations, the deformed shape and total reaction force were then compared to the experimental results and are presented next. 

\begin{figure}
 \centering
  \includegraphics[width=\linewidth]{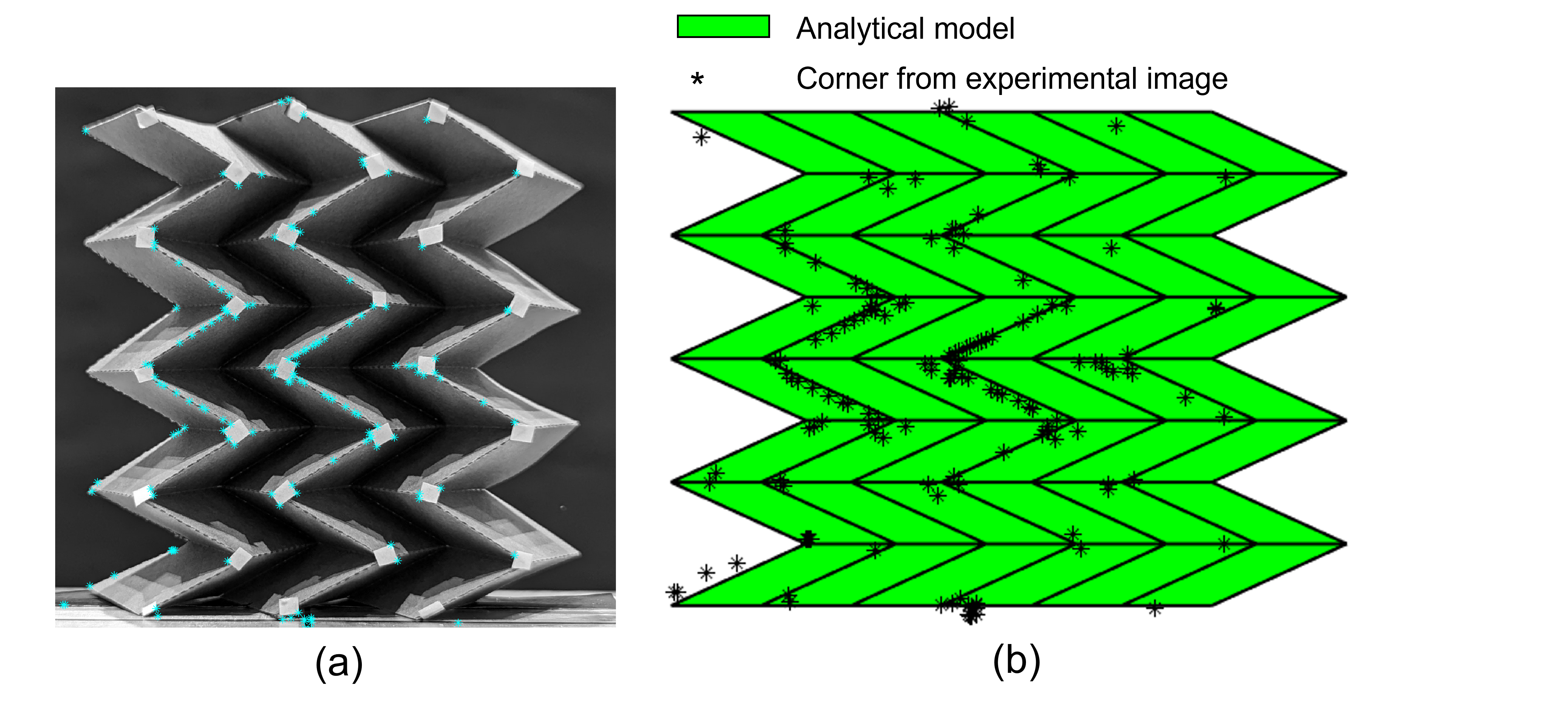}
  \caption{(a) Miura-Ori origami in the undeformed state and detected corners (shown by white `*' marks), (b) Comparison of analytical geometry to experimental geometry in the undeformed configuration }
  \label{fig:mo_undef}
\end{figure}

\begin{figure}
 \centering
  \includegraphics[width=\linewidth]{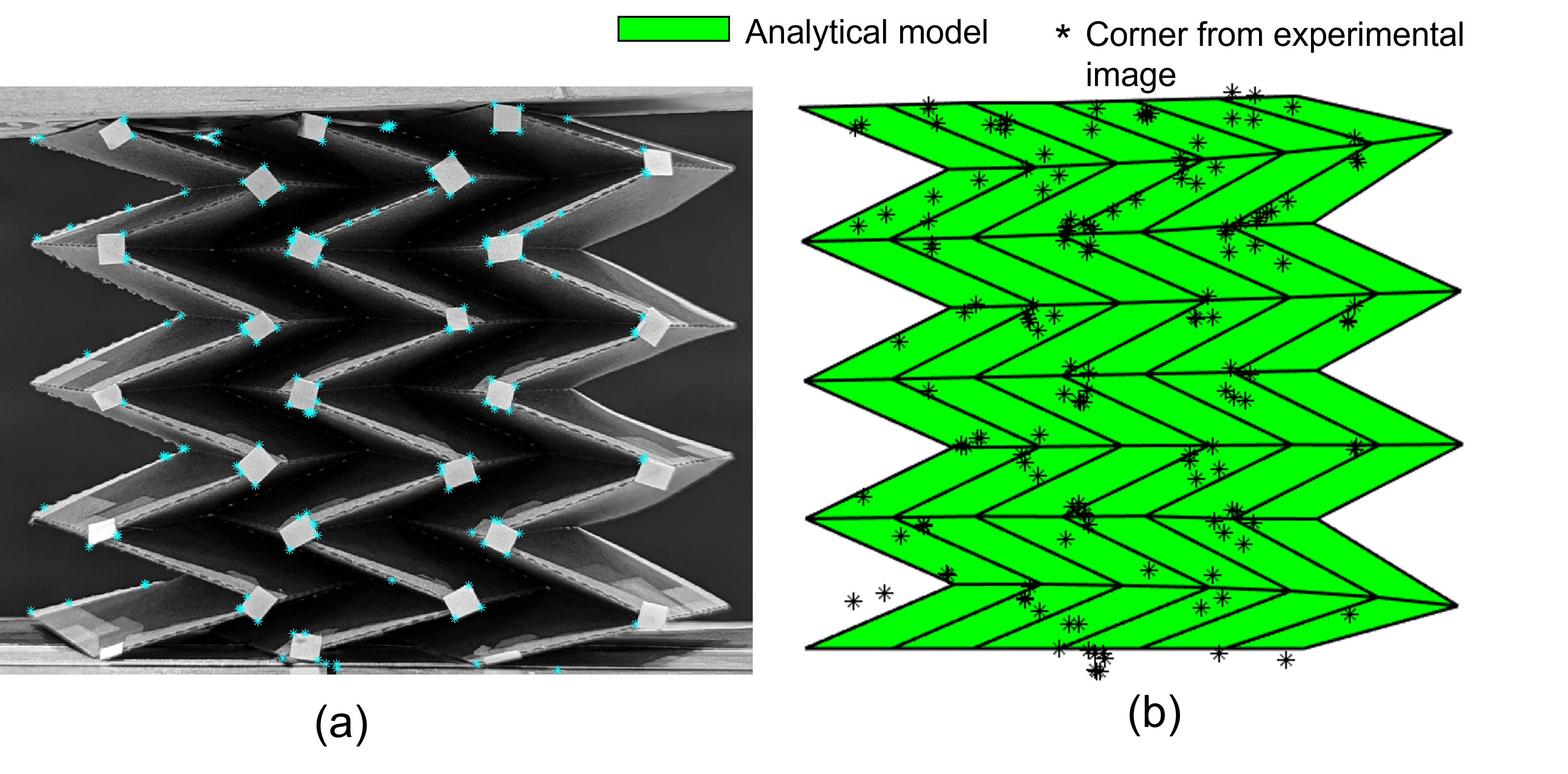}
  \caption{(a) Miura-Ori origami after a loading of 12.75 grams and detected corners (shown by white `*' marks), (b) Comparison of analytical geometry to experimental geometry for a loading of 12.75 grams. }
  \label{fig:mo_l1}
\end{figure}

Figures \ref{fig:mo_undef} and \ref{fig:mo_l1} compares the geometry of the analytical model with the geometry of actual Miura-ori used for experimental investigation. Figure \ref{fig:mo_undef}(a) shows the origami without any applied load. The figure also shows the corners detected using the corner detection algorithm described earlier in the image using white `*' marks. The figure shows that the algorithm detects additional corners along the fold line apart from the vertex locations we are interested in. Figure \ref{fig:mo_undef}(b) shows the analytical model of the origami and overlays it with the calibrated locations of detected corner points from corner detection algorithm. The figure clearly shows that the analytical model captures the location of various corners very well. Further, some corners are also detected due to perforations along the origami folds and they align very well with the edges in the analytical model. Note that for the corners detected on the back face of the sample, the calibration factor will be different from the calibration factor for corners detected on the front face due to the difference in the distances between the camera and the two planes. Thus, these corner points demonstrate the observed differences in Figure \ref{fig:mo_undef} due to differences in calibration factors and because the analytical model is not adjusted for perspective variations. Note that this is true for all the images discussed in the following sections as well.   

\begin{figure}
 \centering
  \includegraphics[width=0.7\linewidth]{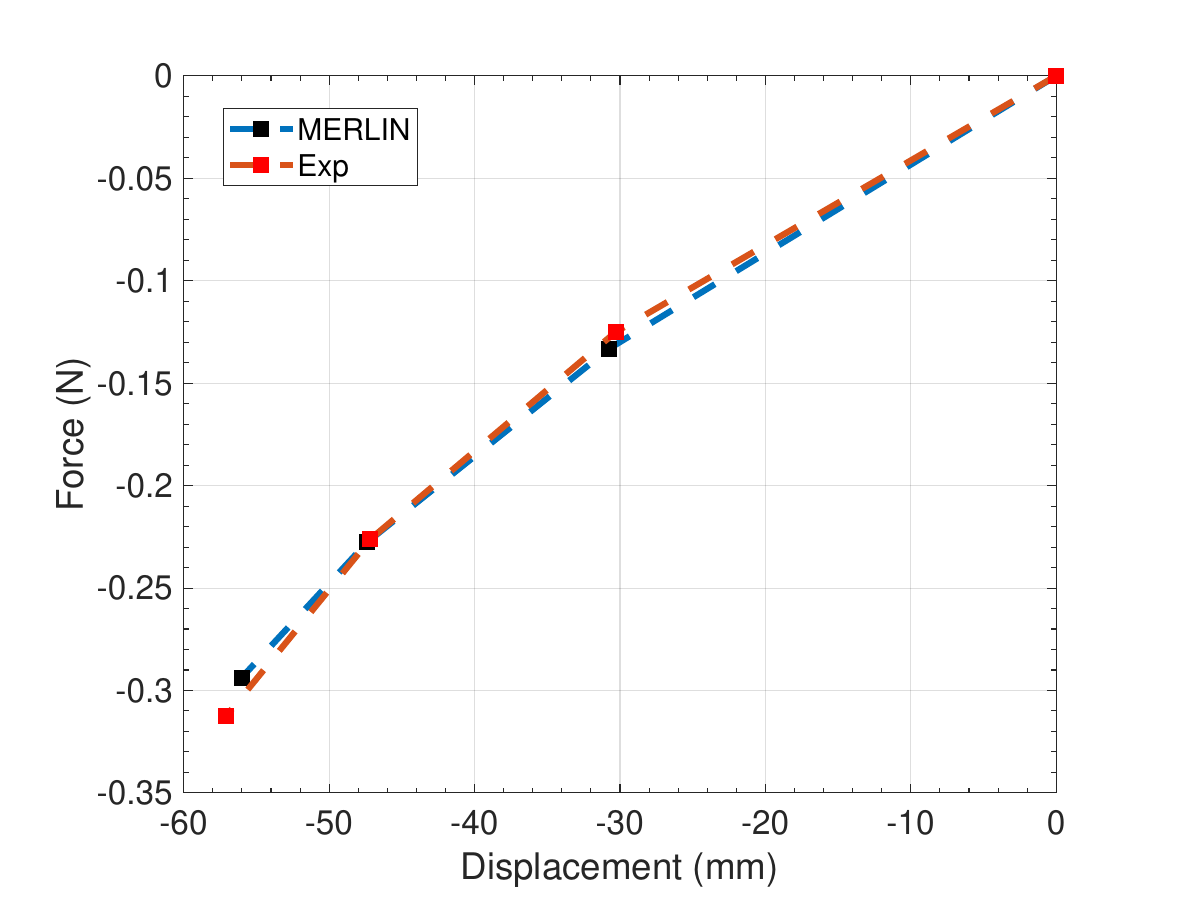}
  \caption{Comparison of total reaction force versus displacement of node 1 results from the experiments to the MERLIN predictions.}
  \label{fig:mo_fd_st}
\end{figure}

Figure \ref{fig:mo_l1}(a) shows the deformed shape of the origami specimen after a loading of 12.75 grams. The displacements of top nodes were computed from the image and imposed on the top nodes in the simulation keeping the bottom nodes fixed. The resulting deformed shape from the simulation is presented in Figure \ref{fig:mo_l1}(b) and is superimposed with the corners detected from the experimental image. The figure demonstrates that the deformed shape from the analytical model agrees well with the deformed shape from the experiment. Further, the total reaction force along the loading direction and corresponding displacement of node 1 in Figure \ref{fig:miura_discrete} is shown in Figure \ref{fig:mo_fd_st}. From the figure, it can be seen that the experimental and numerical forces agree very well.  

Further, the total reaction force versus displacement of node 1 along the loading direction is compared in Figure \ref{fig:mo_fd_st} for experimental and analytical loading and both show a good agreement with each other. The figure also demonstrates the stiffening effect with the tangential stiffness increasing for displacements over $40 \ mm$. This is attributed to the stiffening of the folds as the two surfaces approach closer to each other. It is worth noting that in Figure \ref{fig:mo_fd_st}, the maximum displacement applied was nearly 56 mm while the original height of the sample was nearly 100.48 mm. Thus, the applied strain was over $50\%$, but the model was still able to capture the force-deformation behavior very well as well as the deformation shape of the sample. Further, since force-deformation, as well as the deformed shape, are being captured by MERLIN very well, the stiffness of the Miura-ori sample is being captured by the MERLIN model very well indicating that the dynamic modeling from the analytical model can be validated with the experimental dynamic response of the Miura-ori sample.

\subsection{Dynamic experiments}
The previous section presents the calibration of the MERLIN model for the origami specimen as well as demonstrates that the calibrated model is able to capture the quasi-static behavior of the origami specimen satisfactorily. Thus, we can now proceed to validate the theoretical modeling framework presented in section \ref{sec:dyn_model} with triangle circumcenter mass distribution using experimental results from dynamic testing of the specimen. This section presents the experimental setup and experimental results along with the frequency response function that can be determined using the developed analytical model.

\subsubsection{Experimental setup}
Various dynamic loading scenarios have been used in the structural dynamics literature. For a generic dynamic loading, the response of the system depends both on the system properties (i.e. mass, damping, and stiffness) as well as the loading itself. However, for the impulse loading response and free vibration response, the response only depends on the system properties and therefore they have been used to determine the structural properties. Impulse response refers to the response of the system after the application of initial velocity to the system while the free vibration response refers to the response of the system after the application of initial displacement. By removing the complicating factor of imposed loading, these strategies make it easier to determine the system properties. In this study, the free vibration response of the Miura-ori specimen was used to validate the model.  

To determine the free vibration response of the system, an experimental setup was designed. The main objective of the experimental setup was to impose compressive displacement on the Miura-ori specimen using a panel at the top of the specimen in such a way that the panel is supported at two ends and supports can be removed at the same time swiftly to allow for in-plane excitation and subsequent oscillations. The ensuing response of the specimen can then be recorded using a camera and further processed to determine the response of the specimen. 

\begin{figure}
 \centering
  \includegraphics[width=0.8\linewidth]{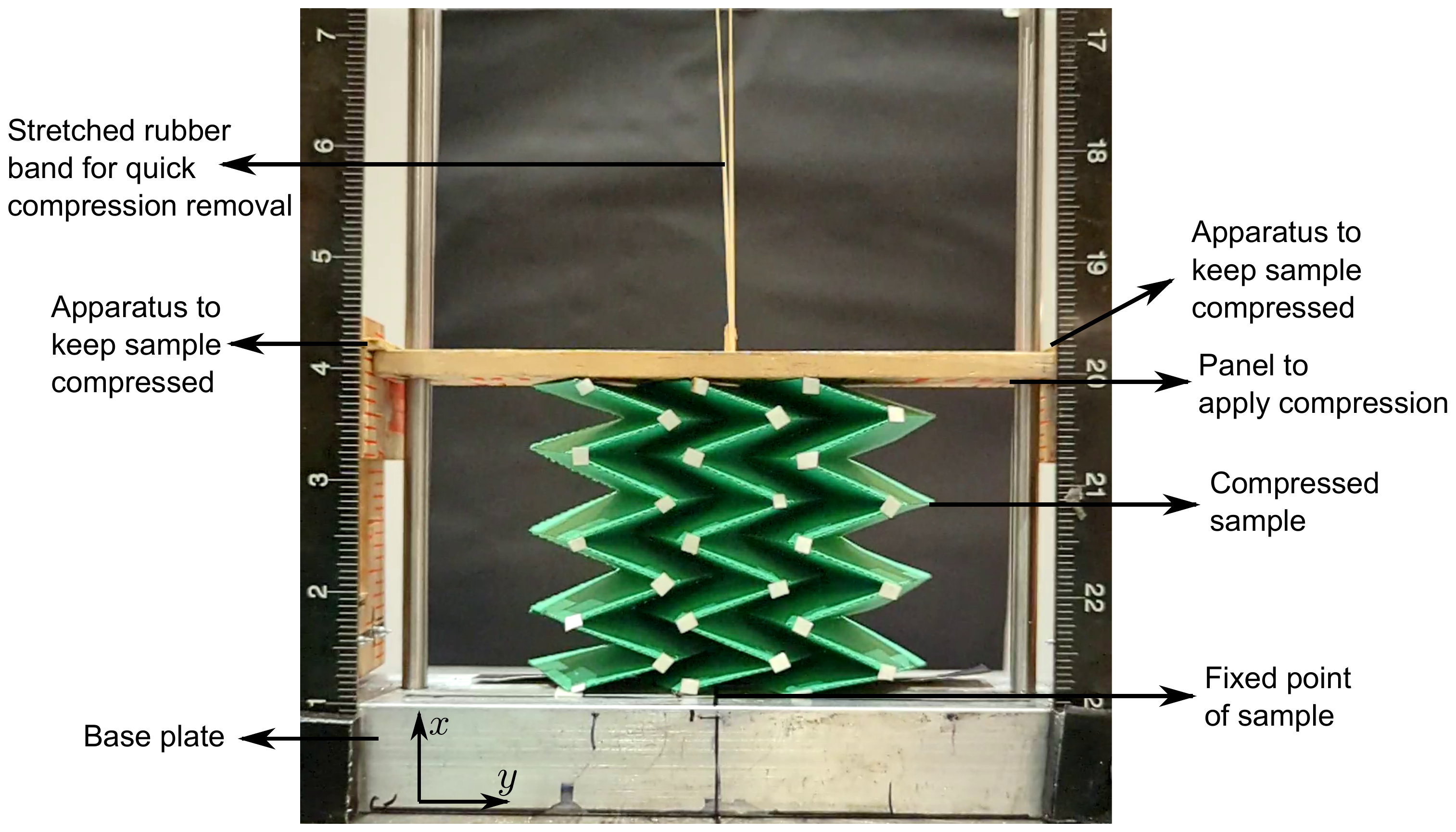}
  \caption{Experimental setup for the dynamic experiments.}
  \label{fig:mo_exp_setup_dyn}
\end{figure}

With the described requirements, an experimental setup was designed and shown in Figure \ref{fig:mo_exp_setup_dyn}. For the dynamic testing, the specimen cannot simply be placed on the base plate as was done for the static case as the specimen will jump after the release of the initial compressive displacement. The ideal support condition to perform the experiment is to fix all the nodes of the bottom surface along the $x$ direction while fixing just the central nodes along the $y$ and $z$ directions. However, this is difficult to achieve practically. It is also not ideal to fix multiple nodes on the bottom surface of the specimen since it introduces unnecessary rigidity to the sample as the response of the Miura-ori specimen is very sensitive to boundary conditions in the lateral directions. Therefore, only the central node of the bottom surface of the specimen was fixed to the base plate using hot glue.  

At the start of the test, the specimen was held compressed using a wooden panel at the top surface as shown in Figure \ref{fig:mo_exp_setup_dyn}. The panel was supported at the two ends to be in this configuration while ensuring it is horizontal. The supports themselves were connected to the base plate using a hinge connection which allowed for their easy movement which then resulted in easy and quick removal of the supports of the panel. Moreover, the panel was also tensioned from a fixed top support using a rubber band. Both these mechanisms helped in the swift release of the panel after the supports retracted, thereby ensuring that the specimen is free to oscillate by itself without the panel imposing any additional motion on the specimen during its release. This is needed because the imposition of additional motion might excite modes other than vertical motion which might not be synchronous with the vertical mode. This would make the estimation of the state of the origami specimen extremely hard. 

The data acquisition for the experiment was performed using a mobile phone camera. Videos were captured at a frame rate of 240 per second which was sufficiently high for the frequencies of oscillation as well as for post-processing using the object tracking algorithm. The videos had a resolution of $2268\times4032$ pixels. The camera was placed in front of the setup with the camera plane being parallel to the plane of the sample. 

\subsubsection{Experimental results}
\begin{figure}
 \centering
  \includegraphics[width=0.8\linewidth]{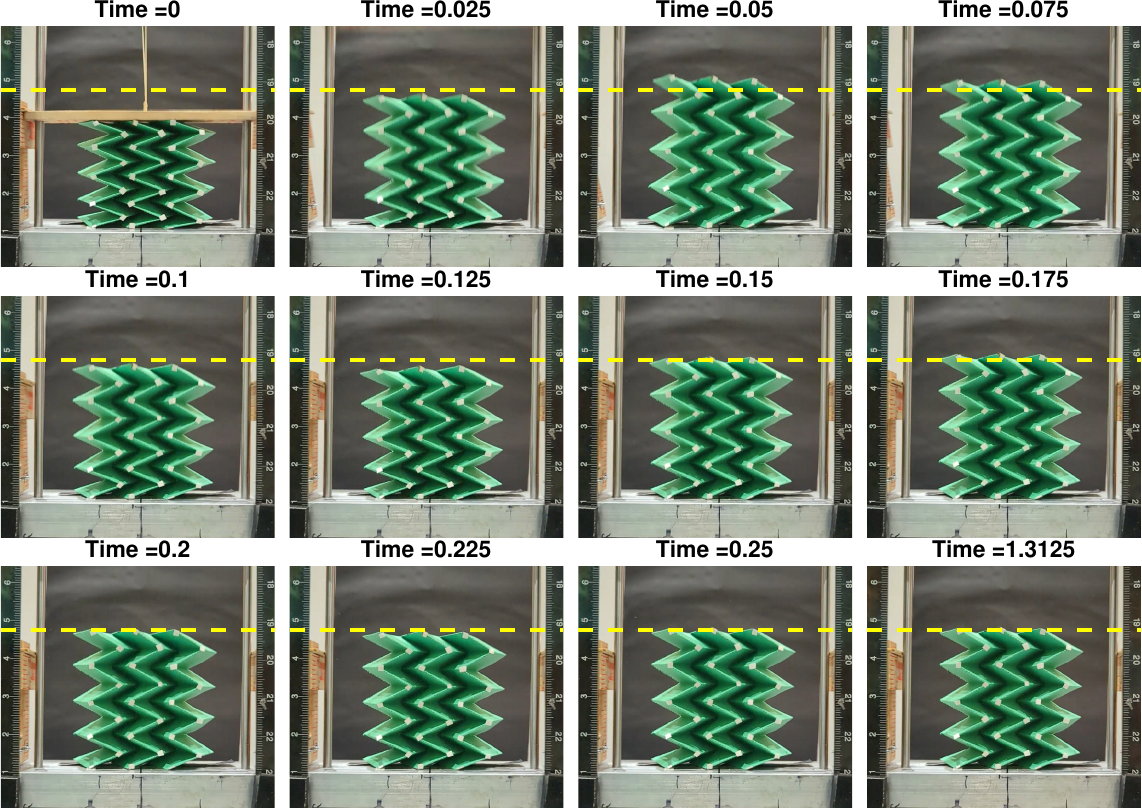}
  \caption{Snapshots of oscillations of Miura-ori sample at different time instances from the video.}
  \label{fig:mo_dyn_exp_snap}
\end{figure}

Using the setup described in the previous section, dynamic experiments were performed. The snapshots of the video obtained from the experiment at various time instances are shown in Figure \ref{fig:mo_dyn_exp_snap}. The snapshots contain a yellow dashed line which helps in visualizing the oscillations.  

\begin{figure}
 \centering
  \includegraphics[width=0.8\linewidth]{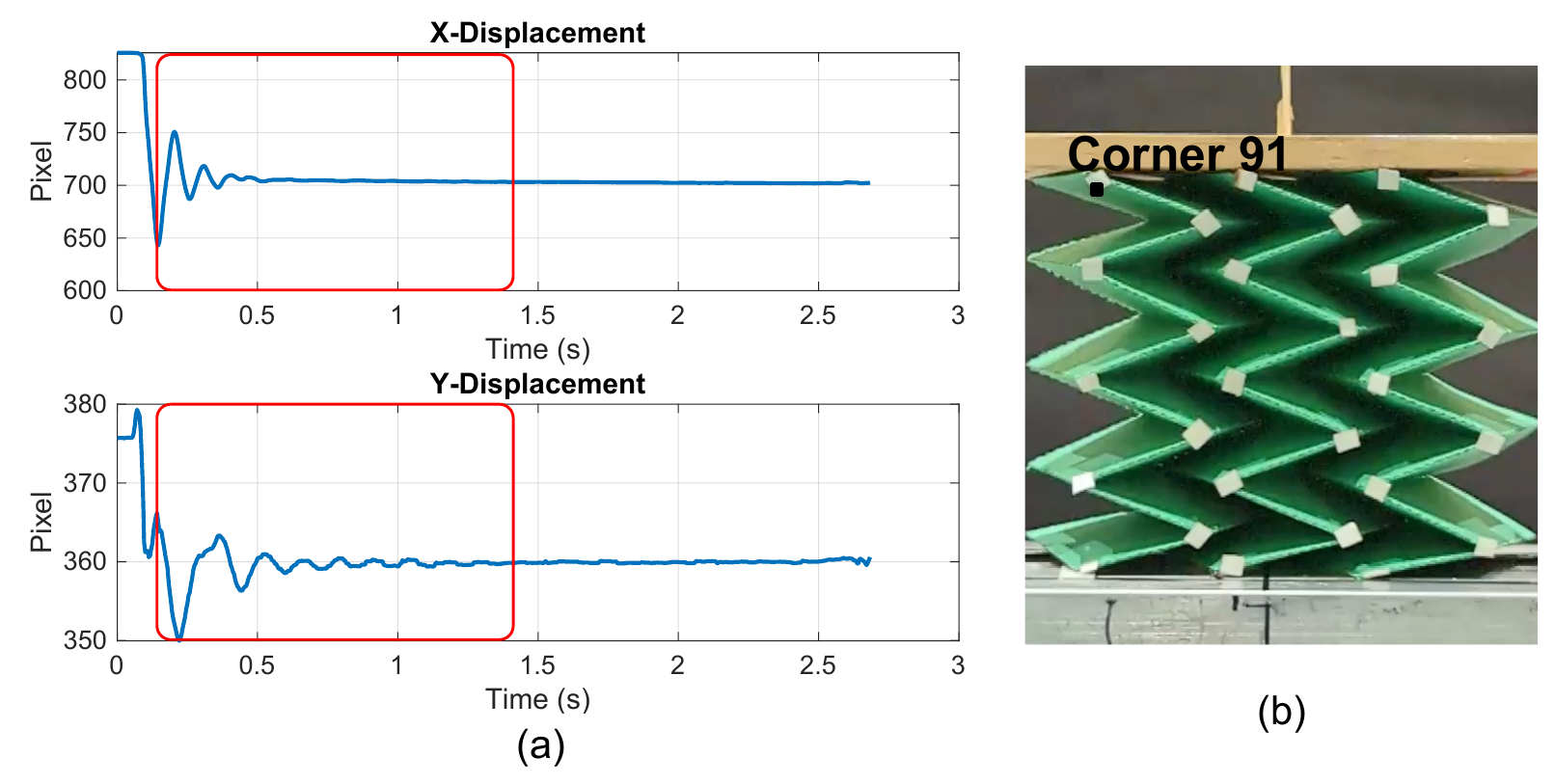}
  \caption{Time history of the location of corner 91. The red box indicates the selected time history. }
  \label{fig:mo_dyn_node_loc}
\end{figure}

Figure \ref{fig:mo_dyn_node_loc}(a) shows the complete time history of the location of a sample corner points \textit{Corner 91}, which is shown in Figure \ref{fig:mo_dyn_node_loc}(b). The time history shows that initially, the location is constant and then when the panel is released we can see a large change in the location of the corner point and then it starts oscillating with decay exhibiting the typical free vibration response characteristics. Note that in the lower row of Figure \ref{fig:mo_dyn_node_loc}(a) the initial spike is due to a minor manual adjustment made just before releasing the panel to center the sample and can be ignored. The figure also shows the selected time history for simulations with the red box. The initial jump is not selected because it is possible that the panel is in contact with a part of this motion with the sample and might influence its motion. At $t=0.14 \ seconds$, the Miura-ori sample has expanded to its maximum, and thus all the nodes will exhibit zero velocity along $x$ direction at this point, and thus, it can be assumed that from this point on the Miura-ori will demonstrate free vibration response.

\begin{figure}
 \centering
  \includegraphics[width=0.6\linewidth]{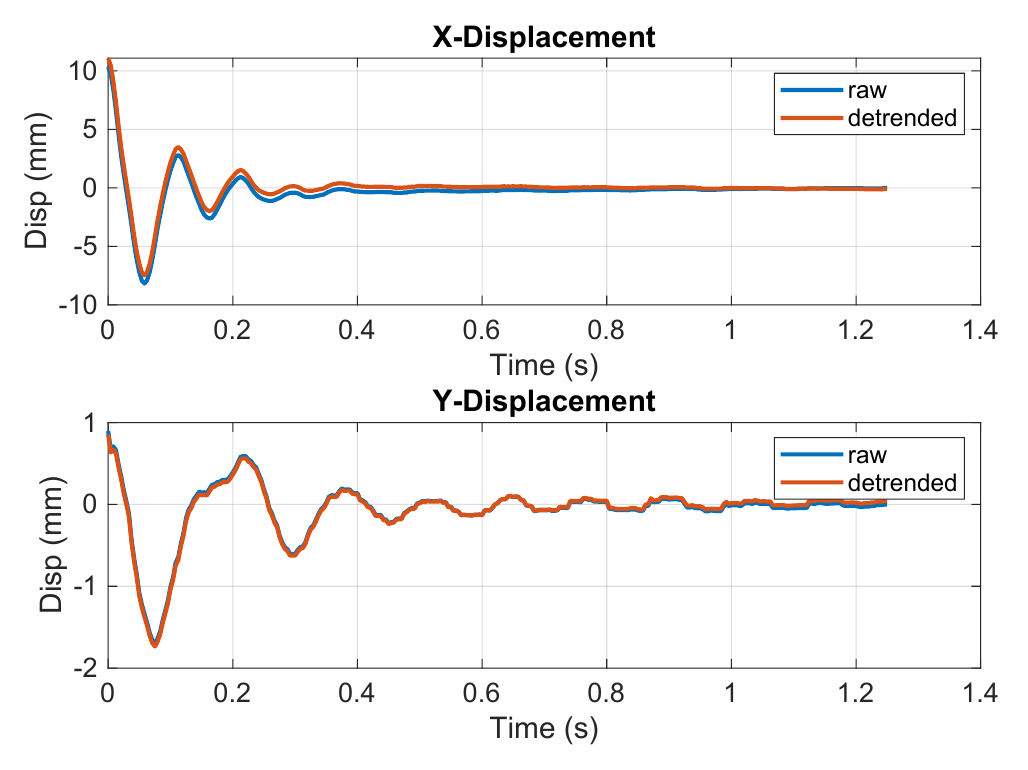}
  \caption{Time history of displacement of corner 91.}
  \label{fig:mo_dyn_node_disp}
\end{figure}

Figure \ref{fig:mo_dyn_node_disp} shows the raw time series of displacement of  \textit{Corner 91} using the selected time history of the location of the corner. The displacement time history was obtained by subtracting the location of the point with the final location of the corner. It should also be noted that the sample tends to slowly expand with time even after vibrations have died down leading to a small linear trend in the time history. This happens due to the release of elastic energy that might have been stored in the hinges during the vibration. This component is not indicative of the system properties, which is the primary objective of this study, and thus was removed by de-trending the time histories. The resulting time history is also shown in Figure \ref{fig:mo_dyn_node_disp}. After this pre-processing step, the time history of all the data-acquisition nodes of the Miura-ori sample was thus ready. 

The frequency of the tensile-compression oscillatory behavior of the origami structure was found to be nearly $9.5 \ Hz$. The damping can be computed from the log-decrement method and using the X-displacement time series of \textit{Corner 91}, it was found that damping for the axial mode was nearly 16\%. 

\subsection{Frequency response function}

\begin{figure}
 \centering
  \includegraphics[width=0.7\linewidth]{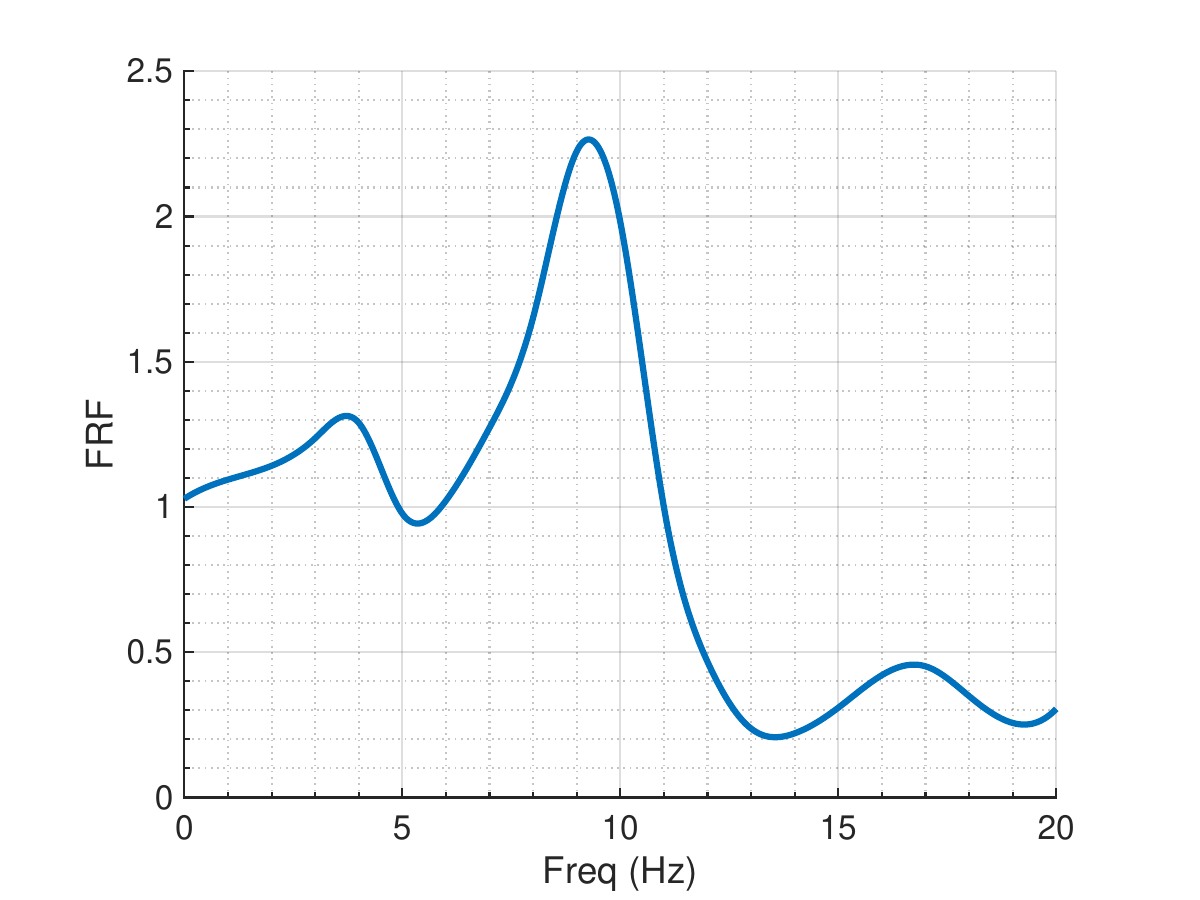}
  \caption{Frequency response function of Miura-ori.}
  \label{fig:mo_frf}
\end{figure}

The frequency response function is defined as the ratio between response (displacement) and excitation (force) as a function of excitation frequency. The frequency response function (FRF) was determined using the displacement response for the described Miura-ori specimen. For the determination of the frequency response function, bottom nodes were constrained along $x$ direction while the top nodes were excited using a force on all the top nodes. The amplitude of force per node was chosen as $0.001 \ N$. The FRF obtained is shown in Figure \ref{fig:mo_frf}. Figure \ref{fig:mo_frf} shows the presence of three dominant modes at $3.8 \ Hz$, $9.3 \ Hz$, and $16.7 \ Hz$. Using the half-power law, the damping ratios can be determined for the second and third modes and it was found that the damping ratio of the second mode was $13.5\%$ and for the third mode it was $9\%$.      

\begin{figure}
 \centering
  \includegraphics[width=0.7\linewidth]{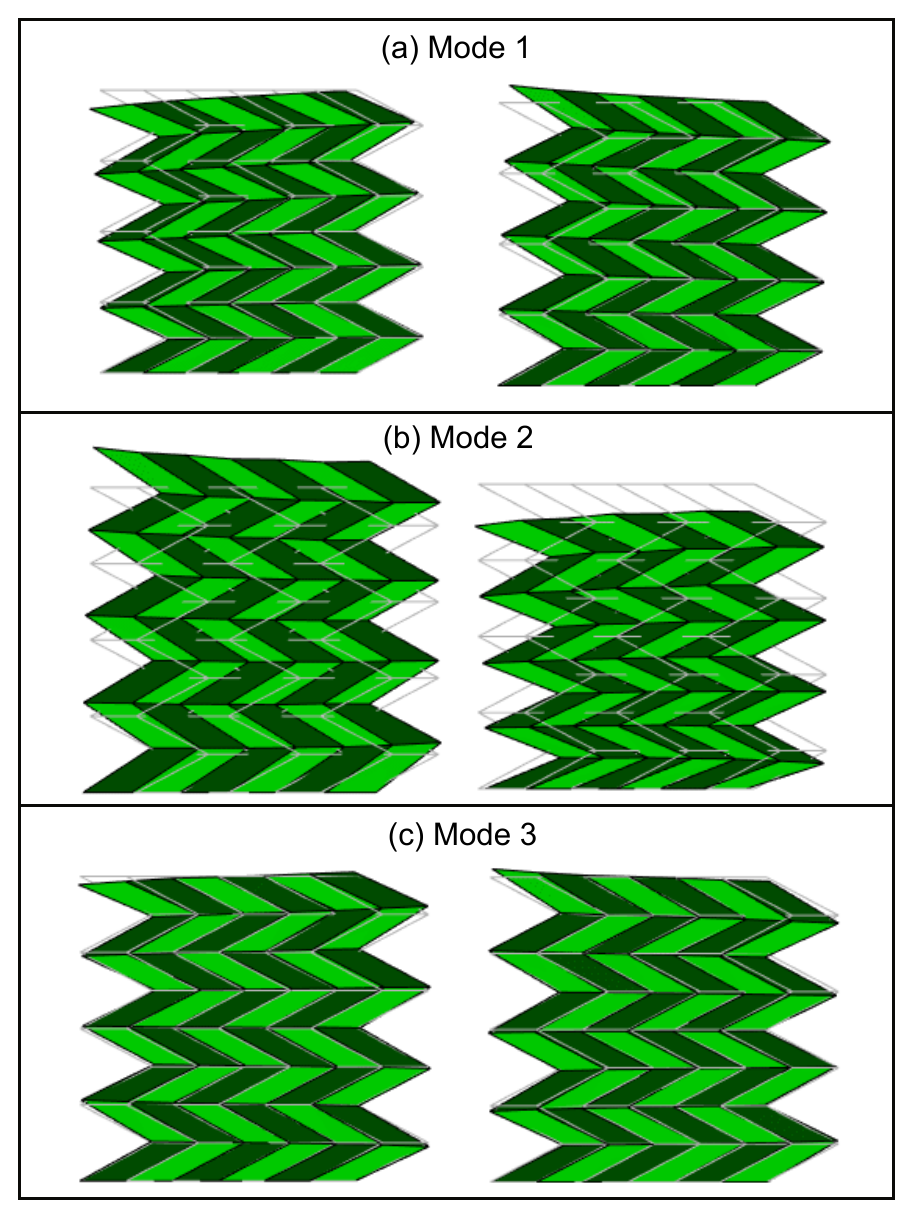}
  \caption{Mode shapes of Miura-ori. The gray lines indicate the undeformed shape of Miura-ori.}
  \label{fig:mo_ms}
\end{figure}

The first mode, i.e. mode corresponding to a frequency of $3.8 \ Hz$ corresponds to a wobbling mode. In this mode, the specimen rocks and can be visualized using the displacement profile as shown in Figure \ref{fig:mo_ms}(a). The second mode, i.e. mode corresponding to a frequency of $9.3 \ Hz$ has the compression-tension mode shape wherein all the top nodes have similar displacements as shown in Figure \ref{fig:mo_ms}(b). The third mode at a frequency of $16.7 \ Hz$ corresponds to the second order wobbling mode as shown in Figure \ref{fig:mo_ms}(c). Thus, using the analytical model, FRFs can be determined which can then be used to obtain parameters of the dynamic behavior of origami structures such as natural frequencies and damping ratios.

\subsection{Analytical results}

Previous sections described the setup and procedure for determining the impulse response of the origami sample as well as the post-processing of the data that was performed. After this analytical simulations were performed using the experimental data to determine the impulse response of the sample analytically and then compare it with the experimental impulse response. To perform the analytical study, we need the displacement and velocity of all the nodes of discretized Miura-ori sample at some instance, which can then be used as initial conditions to perform the dynamic simulations. Referring back to the discussion in the pre-processing state, the easiest time instance for this is when the Miura-ori has reached its maximum tensile deformation after its compression is released as the most components velocity of the nodes will be zero. However, this still poses the challenge of determining the displacement of all the nodes at this time instance. This can be circumvented by a simulation using the displacements of top nodes as the initial deformations and determining the full-field displacement response of the origami nodes. These displacements were then used as initial conditions for the analytical simulations. All the velocity components were taken as zero for the simulations.         

\begin{figure}
 \centering
  \includegraphics[width=0.6\linewidth]{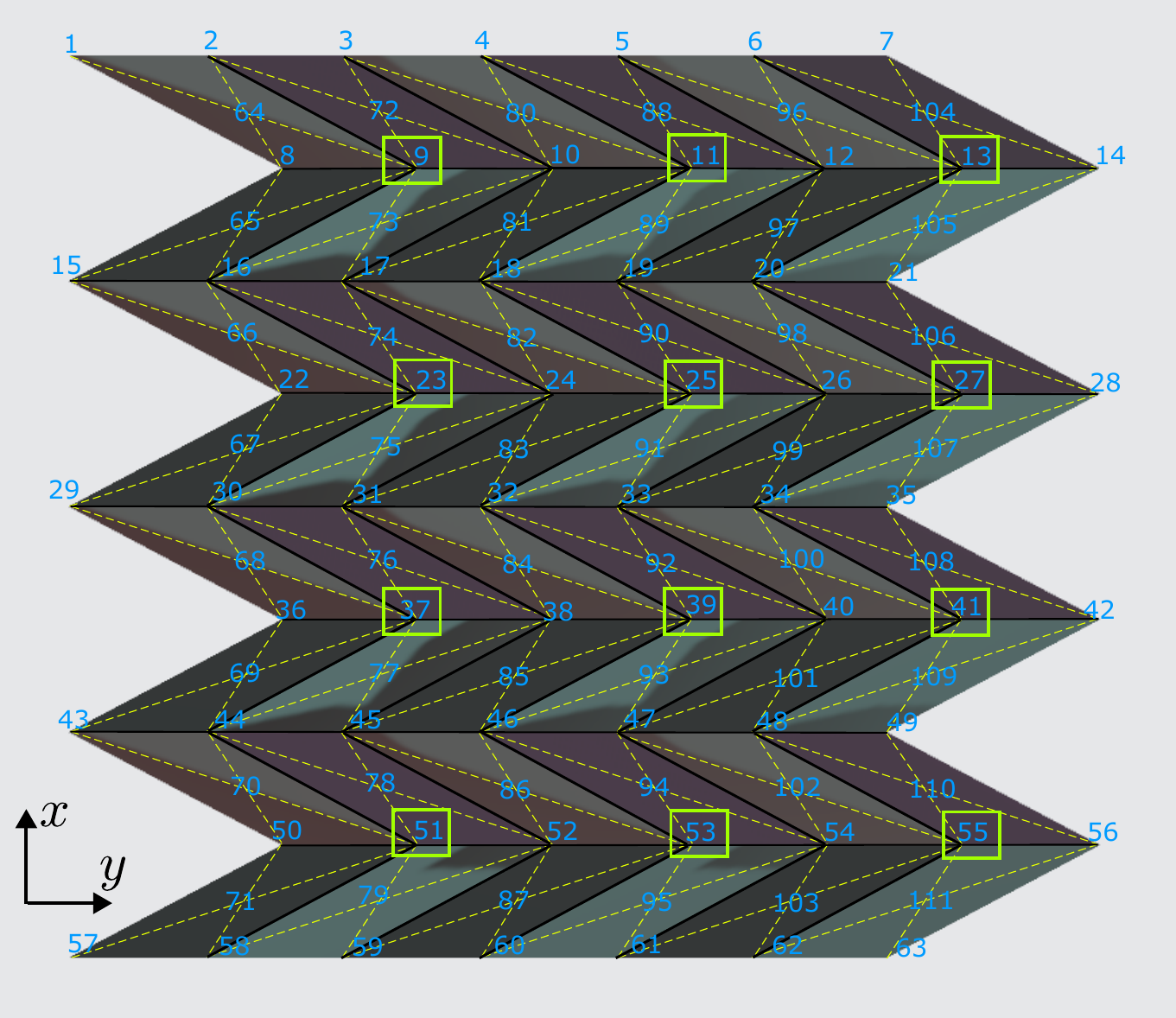}
  \caption{Nodes selected for comparison of their experimental and analytical responses (shown in yellow box).}
  \label{fig:mo_dyn_node_comp}
\end{figure}

The simulations were thus performed using the described initial conditions. Figure \ref{fig:mo_dyn_node_comp} shows the nodes for which the time series data comparison has been presented. The damping matrix was determined by optimizing the difference between the analytical and experimental time histories of $x$ displacement response of just the node $9$. It was assumed that the damping form is proportional damping and thus, $\alpha$ and $\beta$ (where $C = \alpha*M + \beta*K$) were adjusted to determine the optimal parameters.   

\begin{figure}
 \centering
  \includegraphics[width=\linewidth]{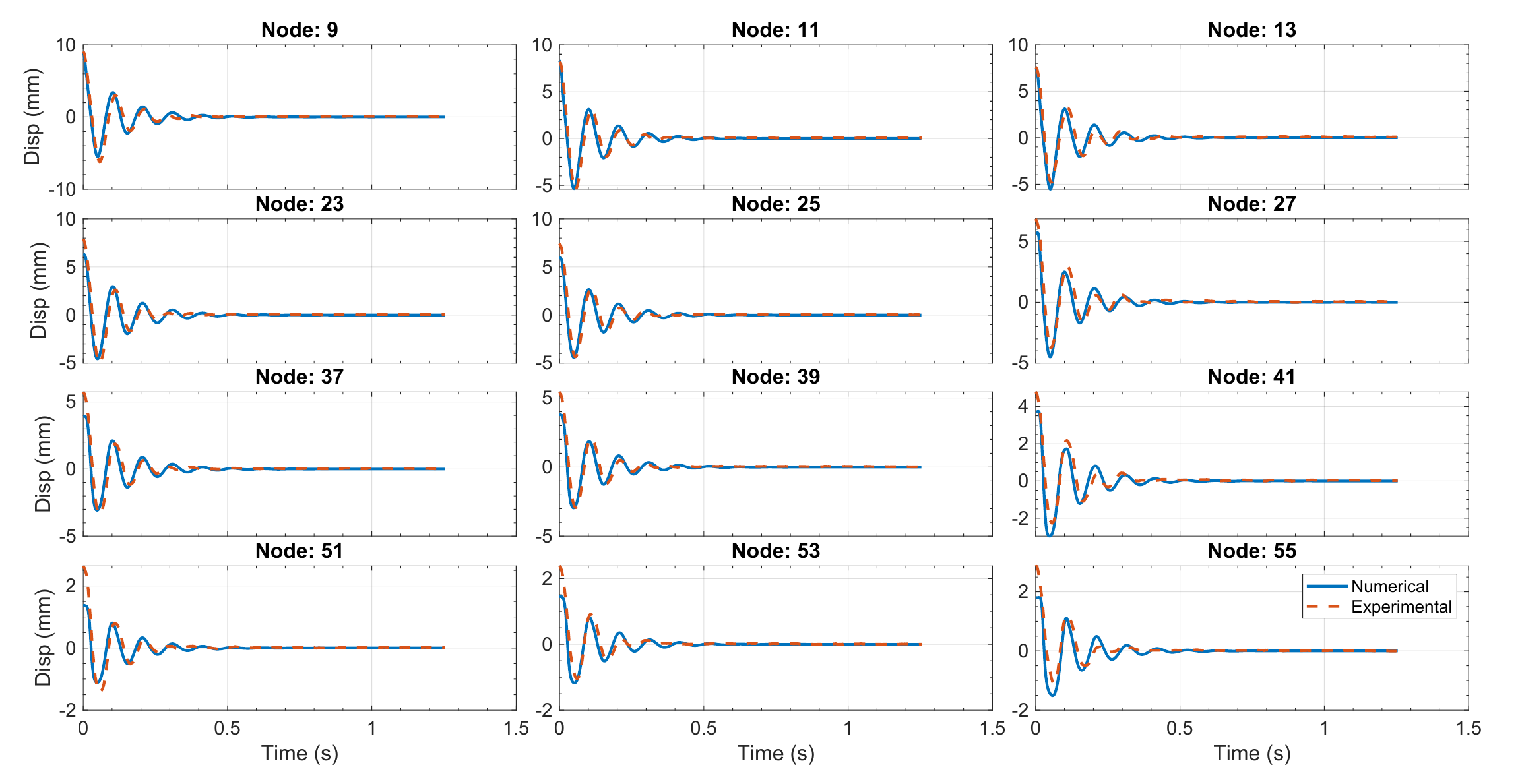}
  \caption{Comparison of experimental and analytical time histories of $x$ displacement of nodes.}
  \label{fig:mo_dyn_thx}
\end{figure}

\begin{figure}
 \centering
  \includegraphics[width=\linewidth]{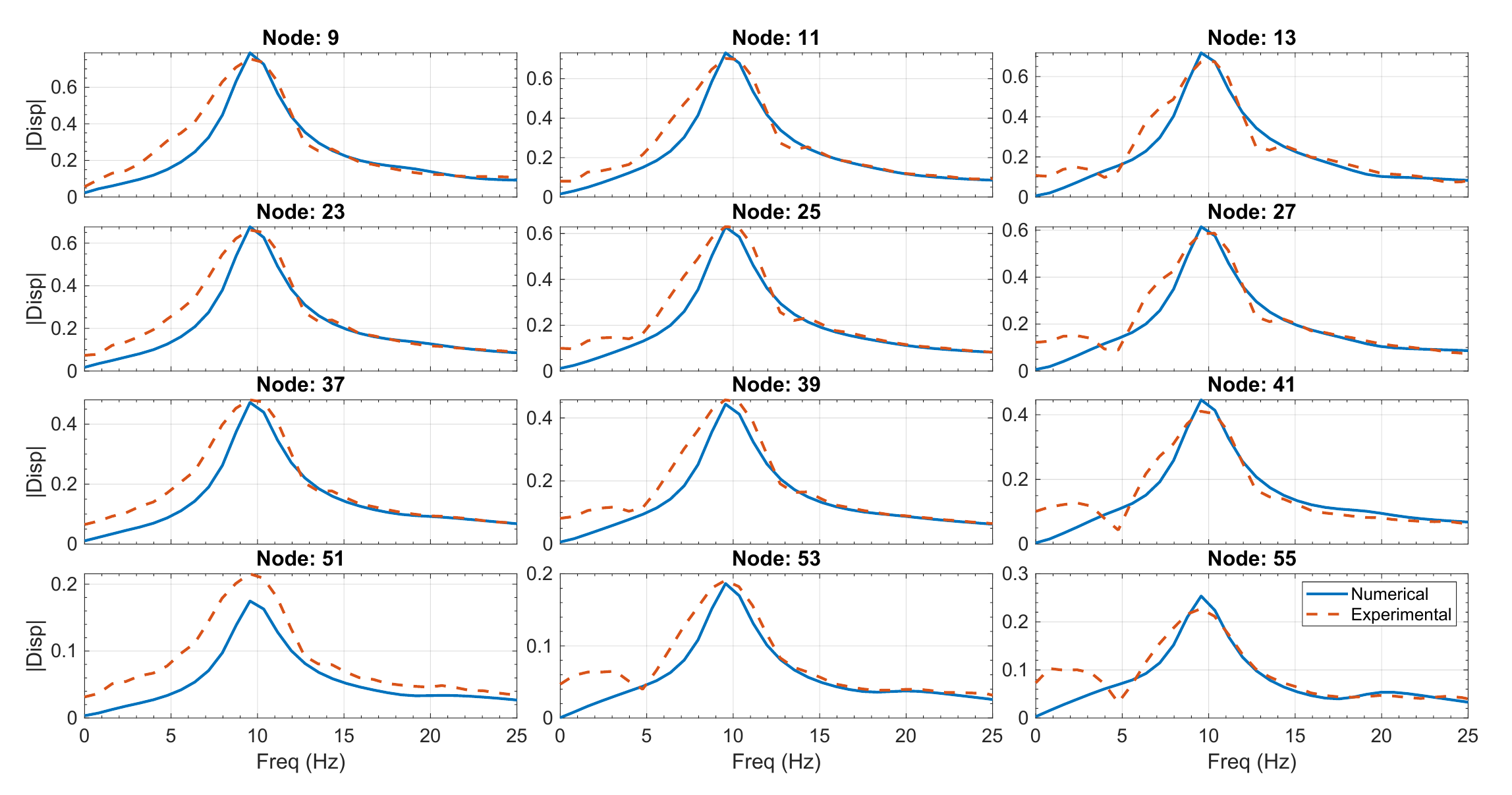}
  \caption{Comparison of the experimental and analytical frequency content of $x$ displacement of nodes.}
  \label{fig:mo_dyn_ftx}
\end{figure}

The $x$ displacement response of all the nodes analytically as well as experimentally is shown in Figure \ref{fig:mo_dyn_thx}. The X-direction response demonstrates the existence of 1 major peak at $9.5 \ Hz$ which corresponds to the axial tension-compression mode (mode 2 of FRF). The figure clearly demonstrates that the analytical model is able to capture the displacements satisfactorily for all the nodes. Note that the displacement of $10 \ mm $ amplitude corresponds to a strain of nearly $10 \% $ for the sample. The corresponding frequency content of the $x$ displacement responses is shown in Figure \ref{fig:mo_dyn_ftx} for all the nodes. The figure also demonstrates the peak frequency as well its corresponding value is accurately captured by the analytical model. This shows that the analytical model developed using MERLIN with \textit{Circumcenter mass distribution} was able to not only capture the frequency but also the dynamics along the $x$ direction fairly well.

\section{Conclusions}
In this study, the dynamic model for modeling the dynamic behavior of origami was formulated. The formulation was based on using MERLIN to capture the stiffness of the origami structures while using the mass distribution formulation to model the mass. Different mass matrix formulations were chosen and implemented. Further, the response from these formulations was compared with FEM for different scenarios covering different loading conditions, geometries, and folding/bending-dominated behavior. It was found that Triangle circumcenter distribution was able to capture most of the dynamics consistently and satisfactorily and can be a good candidate for use. Further, the analytical model for the dynamic behavior of the origami structures was experimentally validated. To perform the validation, the Miura-ori sample was chosen and manufactured. A setup was then manufactured to determine the material properties of the origami structure to calibrate MERLIN. Using the calibrated MERLIN model, it was verified that MERLIN captured the experimental behavior of the sample very well for the case of quasi-static loading. 

After the validation of quasi-static behavior, the focus was then shifted to the dynamic behavior of the sample. The experimental setup was modified to determine the free vibration response behavior of the Miura-ori sample. Using the experimental setup, the Miura-ori sample was compressed and then the compression was swiftly released while the video recorded the behavior of the Miura-ori sample. The video was then processed to compute the displacement of various nodes of the sample and pre-processed to determine its impulse response behavior experimentally. Further, analytical simulations were then performed and compared with the experimental results. The comparison demonstrated that the analytical model was able to capture most of the dynamics in the longitudinal direction. 

\section{Acknowledgements}
Sudheendra Herkal and Prof. Satish Nagarajaiah would like to acknowledge the Science and Engineering Research Board of India (SERB India) funding to Rice University. The authors would also like to thank Albert Daniel Neumann, tech-specialist in Department of Civil and Environmental Engineering, Rice University for all the help with the experiments. We gratefully acknowledge the open-source MERLIN static analysis code \cite{liu2017nonlinear}, which we have updated to implement our proposed dynamic framework.

\clearpage

\bibliographystyle{unsrt} 
\bibliography{refs} 

\end{document}


\maketitle

\section{Case Study 2: Miura-ori X directional loading}
\begin{figure}
 \centering
  \includegraphics[width=0.65\linewidth]{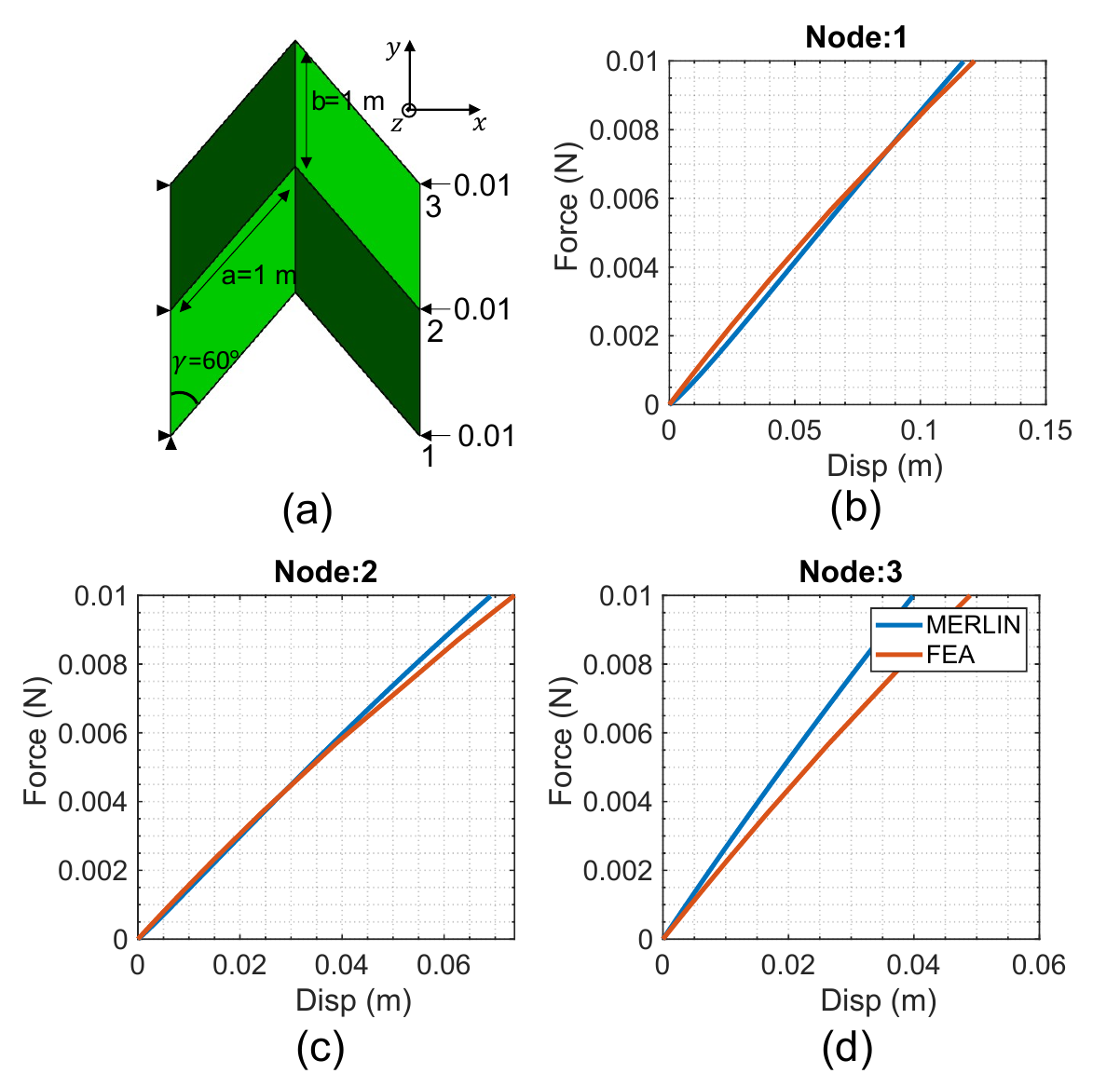}
  \caption{Quasi-static simulation of Miura-Ori under X-direction loading: (a) Geometry, loading and boundary conditions of Miura-ori, (b) Force displacement of node 1, (c) Force displacement of node 2 and (c) Force-displacement of node 3.}
  \label{fig:case_b_qs}
\end{figure}

In the second case study, a more complicated geometry was chosen to test the consistency of these formulations across different geometries. With this consideration, Miura-ori unit cell was chosen as it is one of the most widely used origami geometry for engineering applications. The geometry of Miura-ori as well as its dimensions are shown in Figure \ref{fig:case_b_qs}(a). The panel dimensions of Miura-ori were taken as $1 \ m$ by $1 \ m$ with a thickness of $0.01 \ m$. The dihedral fold angle, $\theta$ was retained at $60 \degree$ and the facet angle $\gamma$ was also taken as $60 \degree$. The geometry of the origami is shown in Figure \ref{fig:case_b_qs}(a). The Young's modulus and Poisson's ratio were retained as $10^6 N/m^2$ and $0.25$ respectively, while the L-scale factor was chosen as $1 \ m$. Similar to the first case study, we perform a quasi-static simulation first to compare the stiffness between FEA and MERLIN2 and then proceed to perform the dynamic simulations. For the static simulations, the loading and the boundary conditions used are shown in Figure \ref{fig:case_b_qs}(a). Figure \ref{fig:case_b_qs} shows that the stiffness of nodes 1 and 2 was well captured but there is a slight difference between the two models for node 3. The reason for the difference can be attributed to the bending deformation near the top panel (near node 3) that was observed due to eccentric loading and higher folding stiffness.

\begin{figure}
 \centering
  \includegraphics[width=0.95\linewidth]{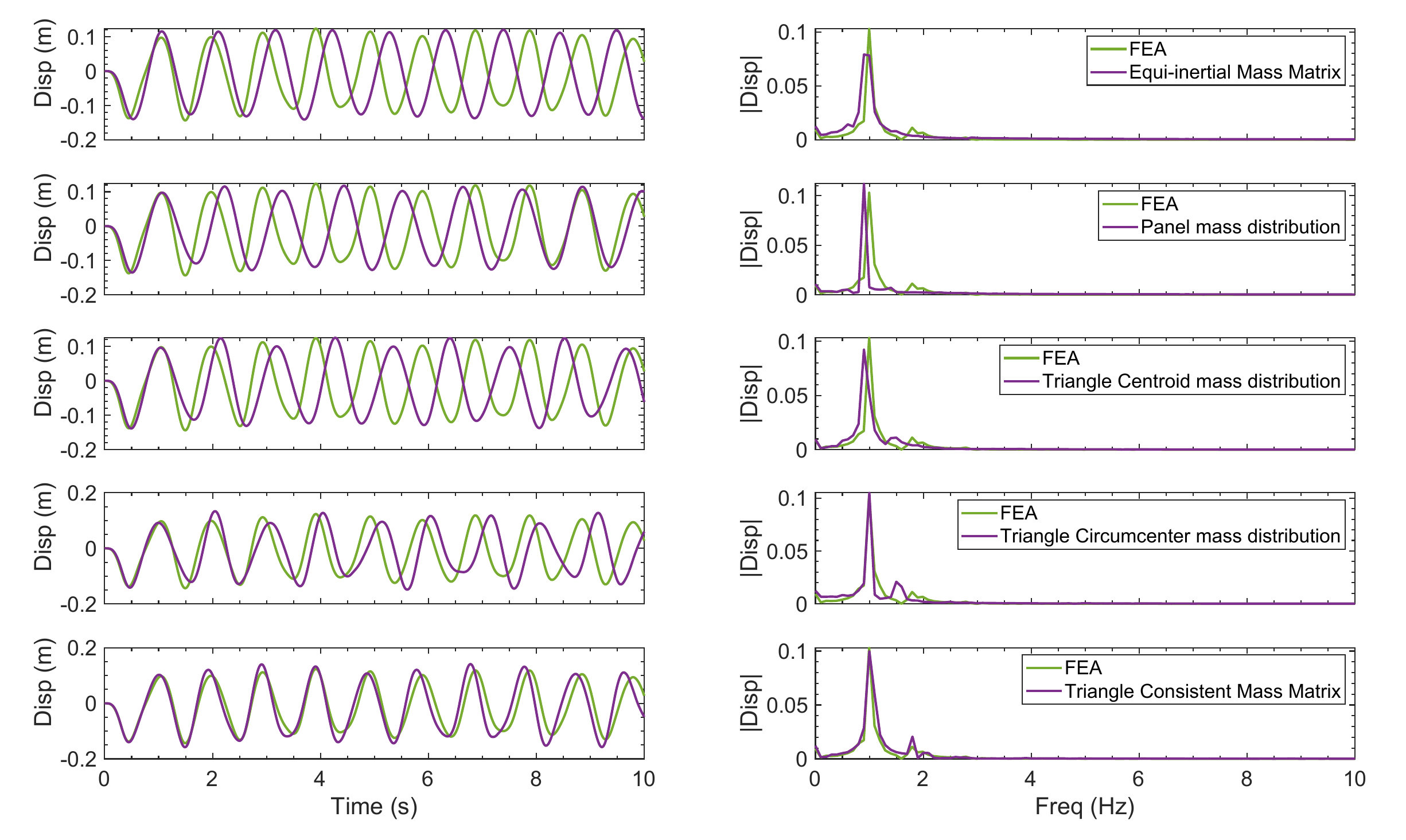}
  \caption{Results from dynamic simulation under X-direction loading of Miura Ori: Left - Time series of displacement of Node 1 with different mass matrices and Right - Corresponding frequency content.}
  \label{fig:case_b_qs1}
\end{figure}

\begin{figure}
 \centering
  \includegraphics[width=0.95\linewidth]{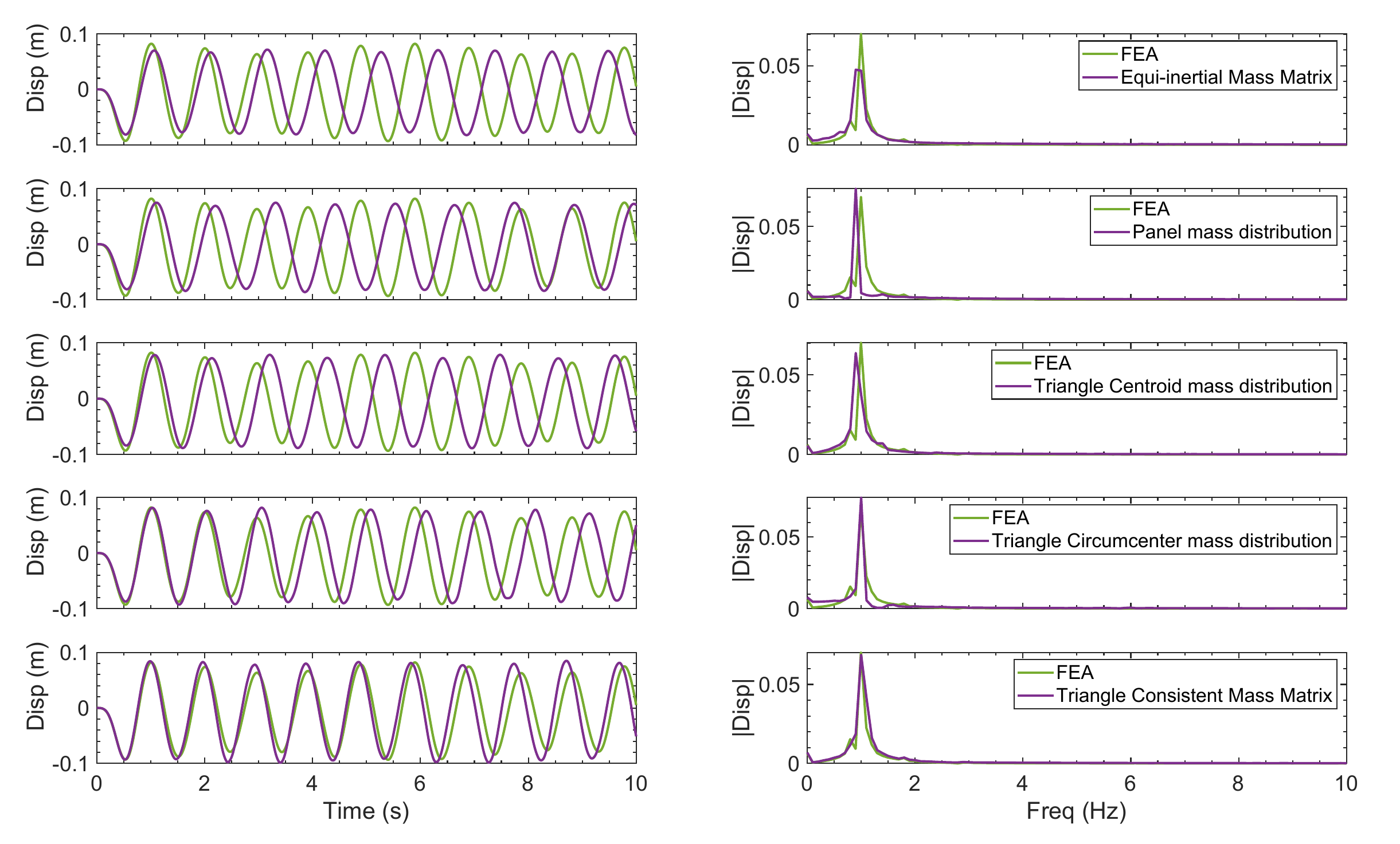}
  \caption{Results from dynamic simulation under X-direction loading of Miura Ori: Left - Time series of displacement of Node 2 with different mass matrices and Right - Corresponding frequency content.}
  \label{fig:case_b_qs2}
\end{figure}

\begin{figure}
 \centering
  \includegraphics[width=0.95\linewidth]{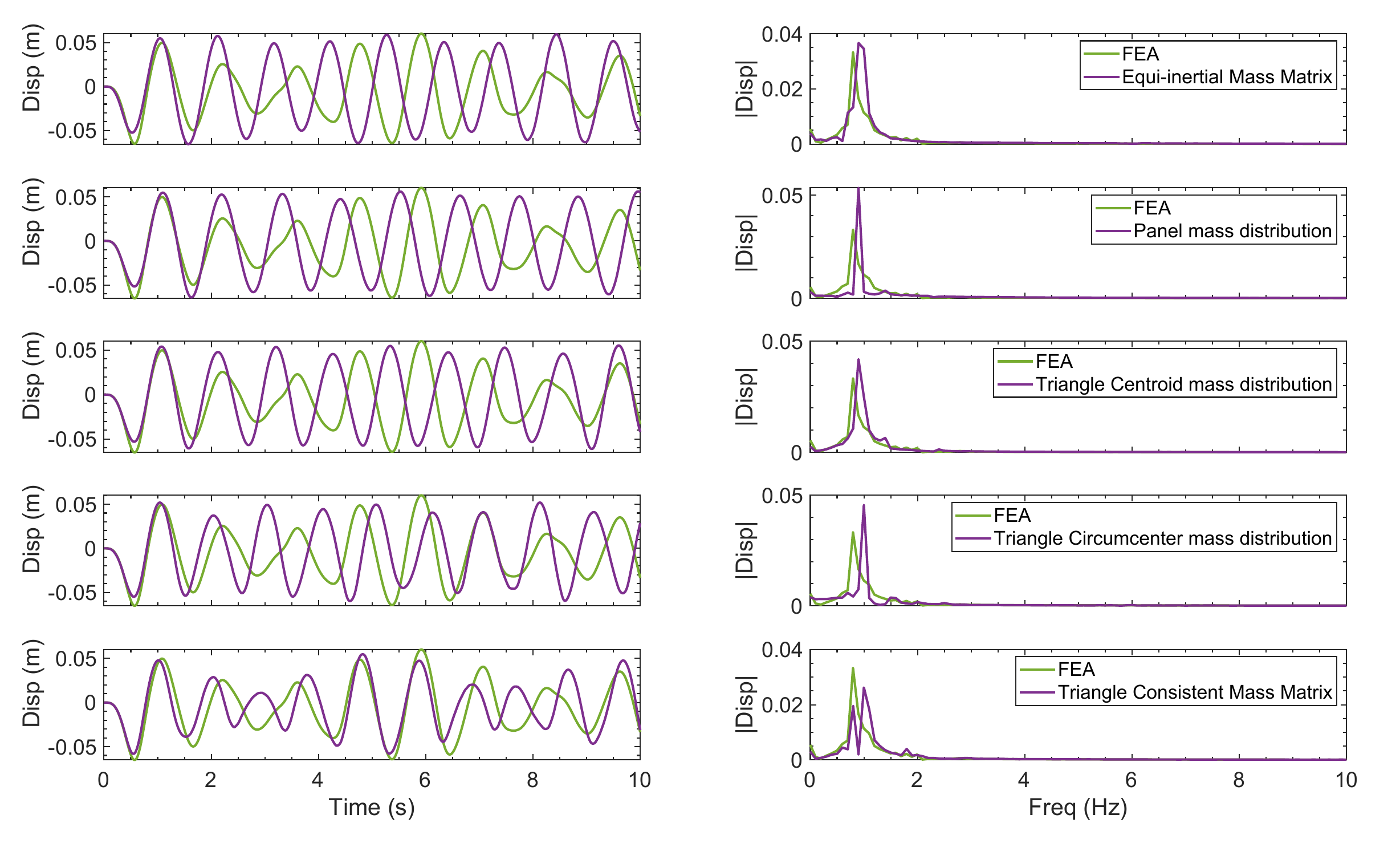}
  \caption{Results from dynamic simulation under X-direction loading of Miura Ori: Left - Time series of displacement of Node 3 with different mass matrices and Right - Corresponding frequency content.}
  \label{fig:case_b_qs3}
\end{figure}

As the difference between the two models is not too significant, we now compare the dynamic response between the FEA model and the MERLIN2 with different mass matrices. Using the dynamic formulation proposed and different mass matrices described earlier, the dynamic simulations were performed with the same material properties, loading and boundary conditions as in the quasi-static case but with the corresponding mass distributions. The loading pattern was very similar to the loading pattern described for simple fold but the amplitude of the loading pattern was 0.01 (instead of 0.001 in the case of simple fold, see Figure 4 in manuscript). The results from the simulations are shown in Figures \ref{fig:case_b_qs1}, \ref{fig:case_b_qs2} and \ref{fig:case_b_qs3}. In the case of nodes 1 and 2 (Figures \ref{fig:case_b_qs1} and \ref{fig:case_b_qs2}), the Triangle consistent mass matrix was able to capture most of the dynamics. It was also able to capture the second peak in the frequency domain located around 1.9 Hz. Triangle circumcenter was the second-best mass matrix formulation again capturing most of the dynamics including the amplitude of the displacements and peak frequency content however it had a narrower peak than FEM which resulted in slight differences in the responses. The other mass matrix formulations were not able to capture the peak frequency accurately leading to the differences in the time series. On the hand, for the third node (Figure \ref{fig:case_b_qs3}), all the formulations were unable to capture the dynamics as accurately as they did for the first and second nodes. This could be because of the fact that the stiffness estimation between MERLIN2 and FEM at the third node was slightly different, which might have led to the differences in the response of this node. Overall, we can see that the current dynamic formulation was able to estimate the dynamics of the system satisfactorily for this case as well with consistent mass matrix and triangle circumcenter mass distribution. 

\section{Case Study 3: Miura-ori Y directional loading}
\begin{figure}
 \centering
  \includegraphics[width=0.65\linewidth]{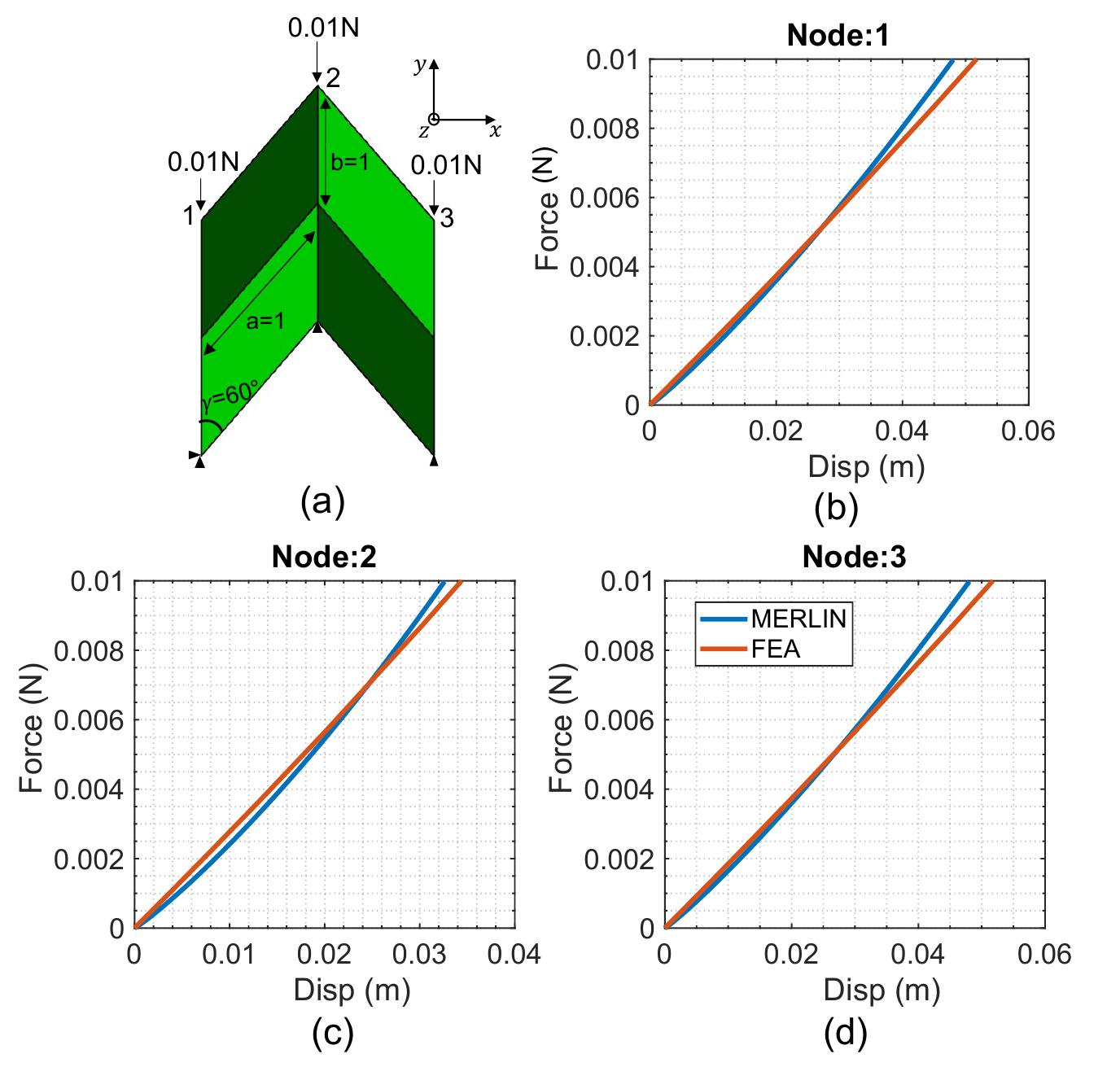}
  \caption{Quasi-static simulation of Miura-Ori under Y-direction loading: (a) Geometry, loading and boundary conditions of Miura-ori, (b) Force displacement of node 1, (c) Force displacement of node 2 and (c) Force-displacement of node 3.}
  \label{fig:case_c_qs}
\end{figure}

The third case we consider is the loading of the Miura-ori unit cell in a different loading direction. This further reinforces the consistency of different mass formulations for various displacement modes in addition to various geometries as well. For this case study, all the geometric and material parameters were retained, i.e. $a = 1 \ m$, $b = 1 \ m$, $\theta = 60 \degree$, $\gamma = 60 \degree$, $E = 10^6 \ N/m^2$ and $ \nu = 0.25$. However, the boundary condition and loading were modified to set the loading along the Y-direction, as shown in Figure \ref{fig:case_c_qs}. The quasi-static simulations were again performed to first check the match between stiffness predicted by MERLIN2 and FEA. Figure \ref{fig:case_c_qs}(b), (c), and (d) show that the stiffness of all the nodes was well captured by MERLIN2. The displacement of Miura-ori in this case was folding-dominated, as the loading was not eccentric. This lets us perform dynamic simulations and compare the efficiency of different mass matrices in capturing the dynamics.   

\begin{figure}
 \centering
  \includegraphics[width=0.95\linewidth]{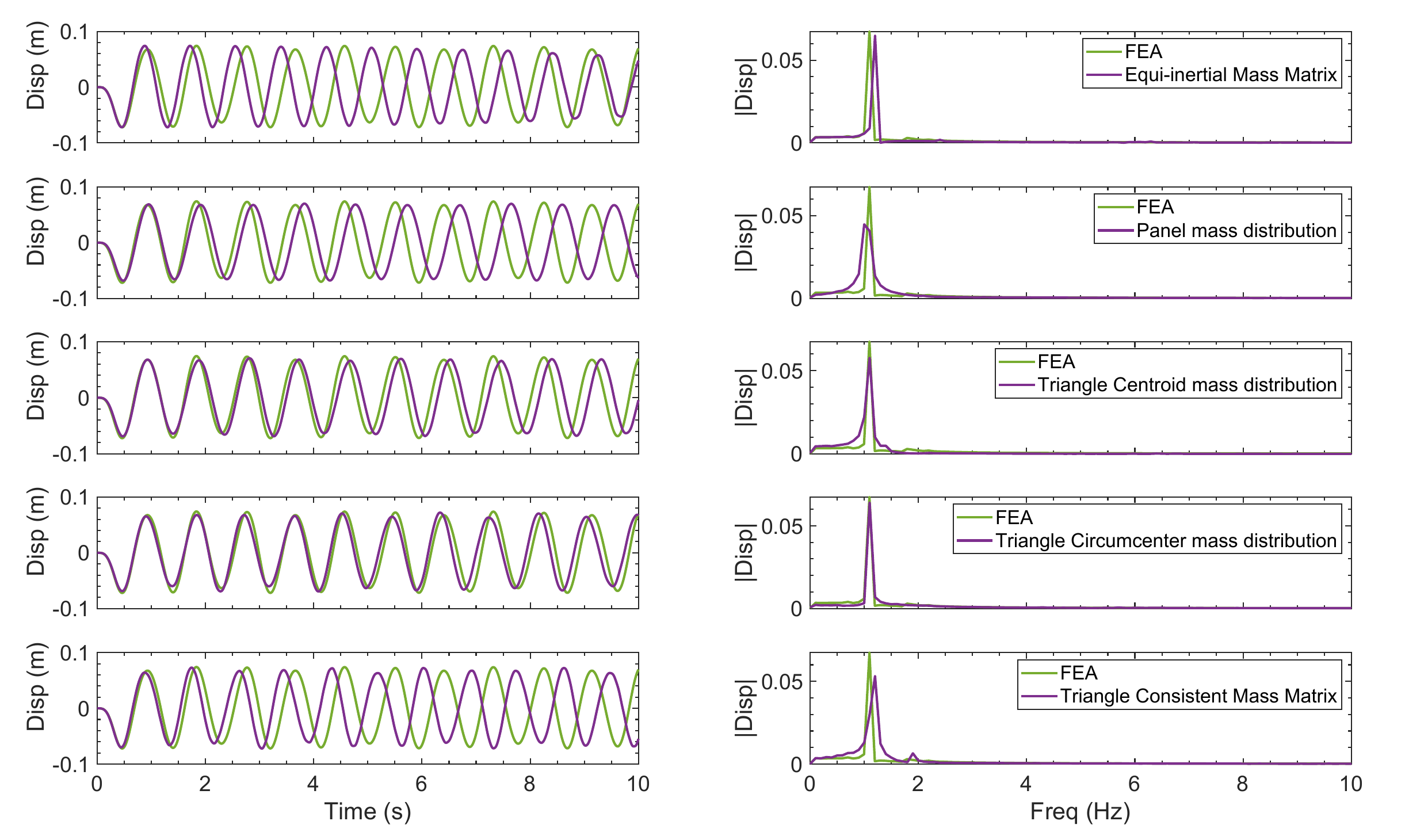}
  \caption{Results from dynamic simulation under Y-direction loading of Miura Ori: Left - Time series of displacement of Node 1 with different mass matrices and Right - Corresponding frequency content.}
  \label{fig:case_c_qs1}
\end{figure}

\begin{figure}
 \centering
  \includegraphics[width=0.95\linewidth]{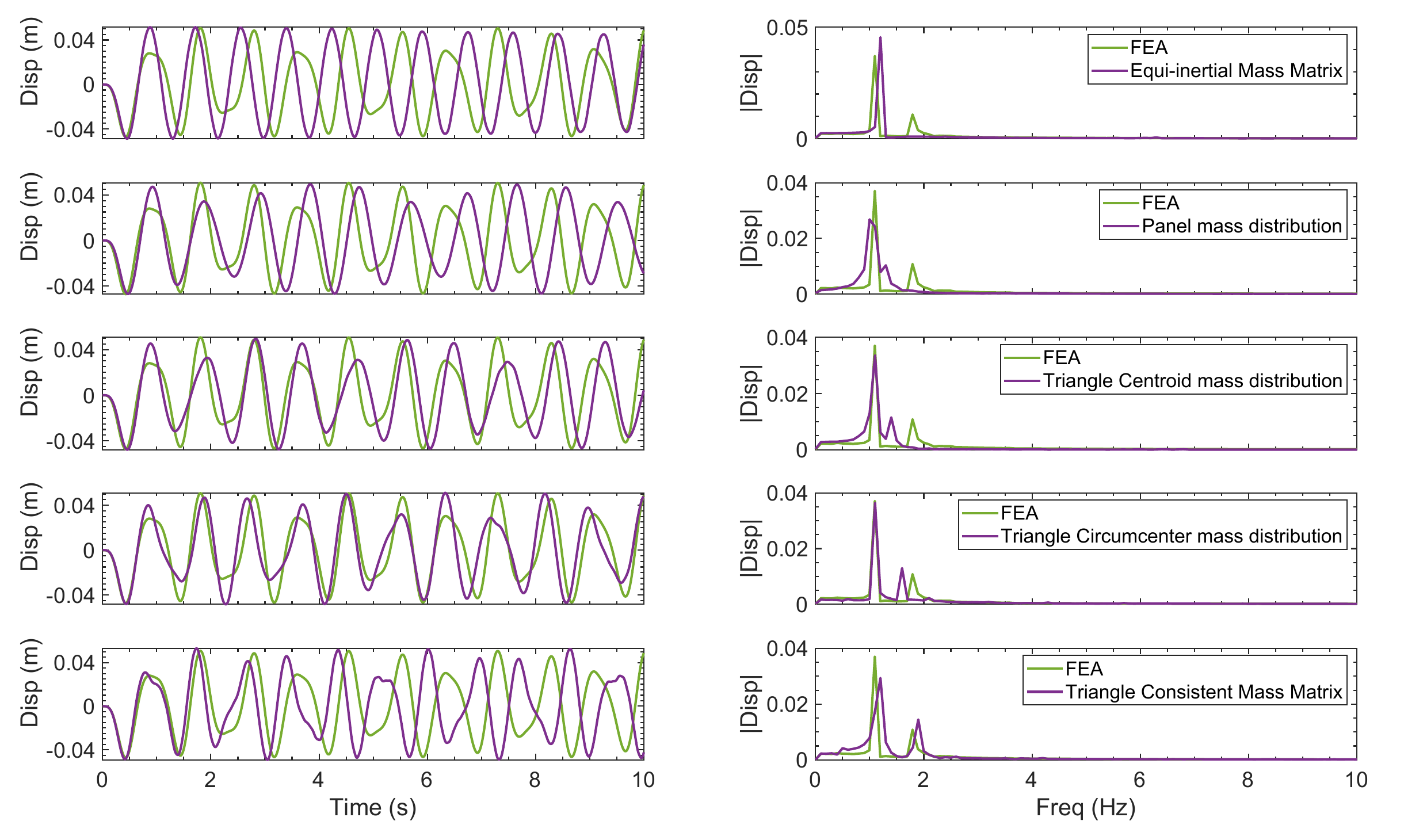}
  \caption{Results from dynamic simulation under Y-direction loading of Miura Ori: Left - Time series of displacement of Node 2 with different mass matrices and Right - Corresponding frequency content.}
  \label{fig:case_c_qs2}
\end{figure}

\begin{figure}
 \centering
  \includegraphics[width=0.95\linewidth]{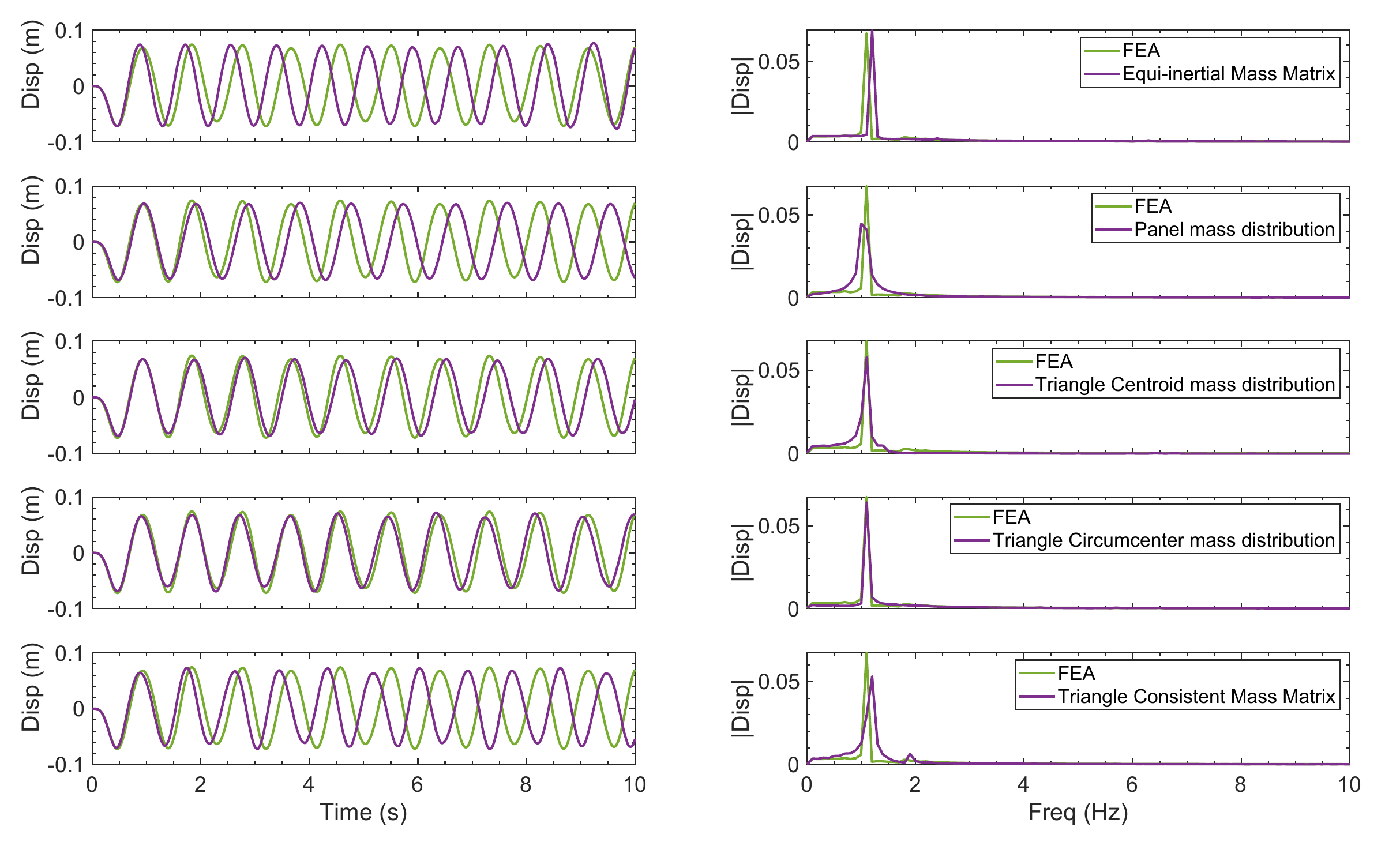}
  \caption{Results from dynamic simulation under Y-direction loading of Miura Ori: Left - Time series of displacement of Node 3 with different mass matrices and Right - Corresponding frequency content.}
  \label{fig:case_c_qs3}
\end{figure}

The results from the dynamic simulations for the three nodes for different mass distribution techniques are shown in Figures \ref{fig:case_c_qs1}, \ref{fig:case_c_qs2} and \ref{fig:case_c_qs3}. From these figures, we can see that the dynamics was nearly captured by both Triangle circumcenter mass distribution and Triangle centroid mass distribution, with a slight error for node 2. Further, we can also see that the error in this estimation was due to a slight difference in the estimation of second frequency peak (around 1.8 Hz for FEA), whereas the first peak was correctly captured. The other mass distribution methods were not able to capture the peak frequency accurately leading to a mismatch between the FEM and their corresponding system responses, although the errors in the peak frequency are not significantly high. Just like in the first two cases, circumcenter mass distribution was able to capture the peak frequency very precisely, leading to fairly accurate results, which shows the consistency of the formulation. 

